\documentclass[a4paper,11pt]{article}
\usepackage{jcappub} 
\usepackage{lineno}

\usepackage{url}
\usepackage{wasysym} 
\usepackage{xcolor}
\usepackage{makecell}
\usepackage{graphicx}
\usepackage{multirow}

\newcommand*{\mkred}[1]{{\color{red}{#1}}}

\newcommand*{\mkgreen}[1]{{\color{green}{#1}}}
\newcommand*{\mkblue}[1]{{\color{blue}{#1}}}
\newcommand*{\mkcyan}[1]{{\color{cyan}{#1}}}

\newcommand{\sv}{\ensuremath{\langle\sigma v\rangle}}

\title{Scrutinizing the impact of the solar modulation on AMS-02 antiproton excess}

\author{Kai-Kai Duan$^a$,}
\author{Xiao Wang$^{a,b}$,}
\author{Wen-Hao Li$^{a,b}$,}
\author{Zhi-Hui Xu$^c$,}
\author{\\Yue-Lin Sming Tsai*$^{a,b}$, and}
\author{Yi-Zhong Fan*$^{a,b}$}
\affiliation{$^a$Key Laboratory of Dark Matter and Space Astronomy, 
Purple Mountain Observatory, Chinese Academy of Sciences, Nanjing 210033, China}
\affiliation{$^b$School of Astronomy and Space Science, University of Science and Technology of China, Hefei, Anhui 230026, China}
\affiliation{$^c$Institute of Modern Physics, Chinese Academy of Sciences, Lanzhou 730000, China}

\emailAdd{smingtsai@pmo.ac.cn, yzfan@pmo.ac.cn}

\abstract{This study examines the impact of solar modulation on the antiproton excess observed by AMS-02, which may indicate dark matter (DM) annihilation. 
We analyze three solar modulation models: the force-field approximation (FFA), a time-, charge-, and rigidity-dependent FFA, and a three-dimensional numerical simulation based on the Parker transport equation. 
Based on AMS-02 latest antiproton data (2025), our results show that the significance of the DM signal is sensitive to the chosen modulation model, with a 2$\sigma$ signal for the FFA (4$\sigma$ if including data from H, He, C, O, B/C, and B/O) and a reduced significance for more complex models. 
We also address systematic uncertainties using two methods: the add-in-quadrature method, which assumes uncorrelated uncertainties between energy bins, and the nuisance parameter method, which treats systematic uncertainties as nuisance parameters during the fitting process. 
Fitted to AMS-02 antiproton data, DM annihilation to the $b\bar{b}$ scenario with three different solar modulation models shows that  
the add-in-quadrature method causes overfitting, whereas the nuisance parameters approach leads to underfitting. 
Statistically, the signal region of the FFA model using the add-in-quadrature method is the most reliable.
This work highlights the need for refined solar modulation models and a better treatment of uncertainties for a conclusive interpretation of the AMS-02 data.}

\begin{document}

\maketitle

\flushbottom

\newpage

\section{Introduction}
\label{sec:intro}

Exploring dark matter (DM) annihilation is key to understanding its nature, although no conclusive signals have been observed. 
However, the excess of antiproton flux observed by the Alpha Magnetic Spectrometer (AMS-02) may be explained by DM annihilation~\cite{2017PhRvL.118s1101C,2017PhRvL.118s1102C,2017PhRvD..96l3010L,2019PhRvD..99j3026C,2022PhRvL.129w1101Z}. 
If DM particles annihilate into Standard Model final states, the antiproton excess suggests a DM mass of around $60-100$ GeV, 
with an annihilation cross-section of roughly $\mathcal{O}(10^{-26})$~cm$^3$s$^{-1}$ for the $b\bar{b}$ final state. 
This suggests considerable interactions between DM and quarks, 
emphasizing the importance of scrutinizing the DM interpretation in light of possible theoretical uncertainties.

Despite the high precision of the antiproton fluxes measured by AMS-02~\cite{2016PhRvL.117i1103A, 2021PhR...894....1A, 2025PhRvL.134e1002A}, theoretical uncertainties associated with cosmic ray (CR) background modeling remain inadequately constrained. 
Consequently, the observed antiproton excess exhibits significant dependence on the adopted background estimation methodology.
Theoretical uncertainties of antiproton fluxes originate from three main sources: (i) CR propagation, (ii) antiproton production cross-sections, and (iii) solar modulation. 
While the first two sources have been extensively investigated for the antiproton excess~(e.g.~\cite{2018JCAP...06..024C,2019PhRvD..99j3026C,2021JCAP...12..037K,2024PhRvD.109d3006L}), 
the impact of different solar modulation remains less explored.
Several studies have addressed these uncertainties from different perspectives. 
Ref.~\cite{2018JCAP...06..024C} uses a Bayesian approach to investigate uncertainties in both the background and DM annihilation components of antiprotons. 
Ref.~\cite{2019PhRvD..99j3026C} revisits uncertainties related to time-, charge-, and energy-dependent solar modulation effects, antiproton production cross-sections, and cosmic-ray propagation, confirming a 4.7$\sigma$ antiproton excess. 
Ref.~\cite{2021JCAP...12..037K} employs neural networks to get rid of nuisance parameters in CR propagation models and perform scans in DM scenarios. 
Ref.~\cite{2024PhRvD.109d3006L} conducts a global Bayesian analysis, reassessing systematic uncertainties in cosmic-ray antiproton flux and finding that the latest AMS-02 data support a purely secondary origin, while setting upper limits on DM annihilation. 
Regarding the complexity of solar modulation in antiproton observations, Ref.~\cite{2023ApJ...953..101A} shows that the modulation effects for antiprotons and protons differ due to their opposing charges, leading to different drift patterns. 
Moreover, Ref.~\cite{2025ApJ...982..103Z} demonstrates that within a modified FFA framework, positron and proton fluxes can be modeled with the same solar modulation parameters, enabling predictions of antiproton fluxes based on electron flux measurements.

The AMS-02 Collaboration released antiproton flux measurements in 2016 and updated them in 2021. 
The initial data provided precise antiproton flux and the antiproton-to-proton flux ratio for CRs within a rigidity range of 1 to 450 GV~\cite{2016PhRvL.117i1103A}. 
The 2021 update extended this range to 525 GV~\cite{2021PhR...894....1A}, confirming the earlier results with enhanced statistical precision.
Recently, the AMS-02 Collaboration released the results over an 11-year Solar cycle of cosmic antiprotons in the rigidity range from 1.0~GV to 41.9~GV~\cite{2025PhRvL.134e1002A}.

To verify the significance of the antiproton excess, we implement three approaches to solar modulation: 
the simple FFA, an extended time-, charge-, and rigidity-dependent FFA formulation, and a comprehensive three-dimensional (3D) numerical simulation based on the Parker transport equation. 
In this work, we include uncertainties from CR propagation and antiproton production, as described in Ref.~\cite{2023FrPhy..1844301M}, to perform a global fit with CR data (except for antiprotons). 
We then use these local interstellar spectra (LIS) based on the posterior distribution to examine the impact of solar modulation on the antiproton excess. 
We start with the simple FFA using an averaged solar modulation potential approximation, neglecting energy losses and drift.
Next, we extend this by considering changes in the solar wind and magnetic field, as discussed in Refs.~\cite{2016PhRvD..93d3016C,2022JCAP...10..051C}. 
Finally, we use a stochastic equation solver~\cite{2022PhRvD.106l3004S} to simulate various Bartel Rotation (BR) cycles of proton data. 
The AMS-02 proton data is used to determine the best-fit parameters for each cycle, which are then applied to propagate antiprotons. 
Using the latest AMS-02 antiproton data, we employ profile likelihood method to assess whether the excess persists and investigate the effects of the three solar modulation models. 
However, the implementation of AMS-02 systematic uncertainties may alter our results, as noted in Ref.~\cite{2020PhRvR...2d3017H}, which simulated the AMS-02 detector to generate a covariance matrix for systematic uncertainties. 
Given the use of public AMS-02 data, unfolding correlated and uncorrelated systematic uncertainties is challenging. 
Thus, in addition to adding statistical and systematic errors in quadrature, we also present results using the nuisance parameter method to account for AMS-02 systematic uncertainties, as in Ref.~\cite{2017PhRvD..95h2007A}.

This paper is structured as follows. In Sec.~\ref{sec:PPC_LIS}, we discuss CR propagation in the Milky Way and the antiproton LIS. Sec.~\ref{sec:modulation} covers solar modulation of CRs, beginning with the FFA in Sec.~\ref{sec:FFA}. 
In Sec.~\ref{sec:model2}, we introduce a time-, charge-, and rigidity-dependent force-field model, 
followed by a 3D numerical simulation based on the Parker transport equation in Sec.~\ref{sec:SDEs}. 
In Sec.~\ref{sec:results}, we present our results, 
including modulation effects and the $\chi^2$ map in Sec.~\ref{sec:chisq_map}, 
methods for addressing systematic uncertainties in Sec.~\ref{sec:sys}, 
possible DM signal regions in Sec.~\ref{sec:excess_region}, 
and best-fit assessment and $95\%$ upper limits in Sec.~\ref{sec:BF}. 
Finally, we summarize our findings and conclusions in Sec.~\ref{sec:conclusion}.

\section{Cosmic Ray propagation and antiproton local interstellar spectra}
\label{sec:PPC_LIS}

\subsection{Cosmic ray propagation in the Milky Way}
\label{sec:PPC}
We employ the propagation model that combines the effects of diffusion and re-acceleration of CRs within the Milky Way~\cite{2007ARNPS..57..285S}.
The propagation equation can be represented as
\begin{align}
\label{eq:propagation}
    \frac{\partial \psi(\vec{r},p,t)}{\partial t}=& 
    \underbrace{q(\vec{r},p,t)}_{\rm source~term}+
    \underbrace{\nabla\cdot(D_{xx}\nabla\psi-\vec{V}\psi)}_{\rm diffusion~term}\nonumber\\
    &+
    \underbrace{\frac{\partial}{\partial p}p^2D_{pp} \frac{\partial}{\partial p} \frac{1}{p^2}\psi}_{\rm re-acceleration~term}- 
    \underbrace{\frac{\partial}{\partial p}\left[ \dot{p}\psi-\frac{p}{3}(\nabla\cdot\vec{V})\psi \right]}_{\rm energy~loss~term}-
    \frac{1}{\tau_f}\psi-\frac{1}{\tau_r}\psi,
\end{align}
where $\psi(\vec{r},p,t)$ is the phase space density distribution of CRs as a function of position $\vec{r}$, momentum $p$ and time $t$. 
The first term is the source of CR $q(\vec{r},p,t)$, while the second and third terms are the diffusion and re-accelerate process, respectively. 
Based on the studies~\cite{2017PhRvD..95h3007Y, 2020JCAP...11..027Y}, 
the diffusion with re-acceleration scenario performs significantly better than the diffusion with convection scenario. 
Therefore, we ignore the convection effect ($\vec{V} = 0$).
The fourth term includes energy losses and adiabatic losses. 
The terms $\tau_f$ and $\tau_r$ represent the fragmentation and radioactive decay time, respectively.

The diffusion process arises from the stochastic scattering of CR particles. 
Given by a broken power-law function of rigidity ($R=p/Ze$), 
the diffusion term in Eq.~\eqref{eq:propagation} can be described by the spatial diffusion coefficient $D_{xx}$, 
\begin{align}
D_{xx}=\beta_{\rm CR}^{\eta} D_0 
\begin{cases}
\left(\frac{R}{R_{b}}\right)^{\delta_1}, \text{for } R < R_{b}, \\
\left(\frac{R}{R_{b}}\right)^{\delta_2}, \text{for } R > R_{b},
\end{cases}
\end{align}
where $\beta_{\rm CR}$ is the velocity of the CR particle in unit of light speed, $\eta$ is a phenomenological parameter to modify the velocity dependence at low energies, $D_0$ is a normalization constant, $R_{b}$ is the broken rigidity, $\delta_1$ and $\delta_2$ are the spectral indexes below and above than the broken rigidity, respectively.

Moreover, the re-acceleration process, driven by the second Fermi acceleration mechanism, occurs when CR particles interact with irregular magnetic fields in turbulent interstellar gas. This can be viewed as momentum space diffusion, described by the momentum diffusion coefficient $D_{pp}$ as
\begin{equation}
    D_{pp} = \frac{4p^2v_A^2}{3\delta_1(4-\delta_1^2)(4-\delta_1) D_{xx}w},
\end{equation}
where $v_A$ is the Alfvén velocity and $w$ is the ratio of magnetohydrodynamic wave energy density to magnetic field energy density. We can set $w=1$ since it can be subsumed in $v_A$.

Finally, the spatial distribution of sources of CRs can be parameterized by
\begin{equation}
f(r, z) = \left(\frac{r}{r_\odot}\right)^\alpha 
\exp \left[-\frac{\beta (r - r_{\odot} )}{r_\odot}\right] 
\exp \left(-\frac{|z|}{z_s}\right),
\end{equation}
where $r_{\odot}= 8.5$~kpc is the distance from the solar system to the Galactic Center, 
$z_s = 0.2$ kpc is the scale width of the vertical extension of source, and 
the two indices are $\alpha = 1.25$ and $\beta = 3.56$~\citep{2011ApJ...729..106T}.

\begin{table}[]
\centering
\resizebox{\textwidth}{!}{%
\begin{tabular}{|ccccc|}
\hline
\multicolumn{1}{|c|}{Parameter} & \multicolumn{1}{c|}{Unit} & \multicolumn{1}{c|}{Description} & \multicolumn{1}{c|}{Prior Range} & Best fit \\ \hline \hline
\multicolumn{5}{|c|}{Propagation parameters} \\ \hline
\multicolumn{1}{|c|}{$D_0$} & \multicolumn{1}{c|}{$10^{28}$ cm$^2$/s} & \multicolumn{1}{c|}{Diffusion coefhicient} & \multicolumn{1}{c|}{(0.2, 16)} & 4.13 \\
\multicolumn{1}{|c|}{$\delta_1$} & \multicolumn{1}{c|}{---} & \multicolumn{1}{c|}{First difusion coeficient rigidity power} & \multicolumn{1}{c|}{(0.01, 1)} & 0.45 \\
\multicolumn{1}{|c|}{$R_{b}$} & \multicolumn{1}{c|}{GV} & \multicolumn{1}{c|}{Broken rigidity} & \multicolumn{1}{c|}{(0.001, 1000)} & 336.03 \\
\multicolumn{1}{|c|}{$\delta_2$} & \multicolumn{1}{c|}{---} & \multicolumn{1}{c|}{Second difusion coeficient rigidity power} & \multicolumn{1}{c|}{(0.01, 1)} & 0.15 \\
\multicolumn{1}{|c|}{$\eta$} & \multicolumn{1}{c|}{---} & \multicolumn{1}{c|}{Diffusion coefcient velocity power} & \multicolumn{1}{c|}{(-5, 5)} & -0.59 \\
\multicolumn{1}{|c|}{$z$} & \multicolumn{1}{c|}{kpc} & \multicolumn{1}{c|}{Height of diffusion zone} & \multicolumn{1}{c|}{(0.2, 16)} & 4.78 \\
\multicolumn{1}{|c|}{$v_{\rm A}$} & \multicolumn{1}{c|}{km/s} & \multicolumn{1}{c|}{Alfven speed} & \multicolumn{1}{c|}{(3, 60)} & 19.15 \\ \hline \hline
\multicolumn{5}{|c|}{Source parameters} \\ \hline
\multicolumn{1}{|c|}{$\lg Q_i^j$} & \multicolumn{1}{c|}{  $\lg {({\rm s}^{-1}{\rm cm}^{-3}{\rm MV}^{-1}})$  } & 
\multicolumn{1}{c|}{
\begin{tabular}[c]{@{}c@{}}
   Injection spectrum of CR. 
   \\If $j=\{p,~{\rm He}\}$, $i$ runs from 1 to 10. Otherwise, $i$ runs from 1 to 6.    
\end{tabular}} & \multicolumn{1}{c|}{(-40, -12)} & --- \\ \hline \hline
\multicolumn{5}{|c|}{Other parameters} \\ \hline
\multicolumn{1}{|c|}{$\Phi_{p}$} 
& \multicolumn{1}{c|}{$10^{-9}$ cm$^{-2}\cdot$ sr$^{-1}\cdot$ s$^{-1}\cdot$ MeV$^{-1}$} & \multicolumn{1}{c|}{Flux of protons at normalization energy 100 GeV} & \multicolumn{1}{c|}{(0.02, 15)} & --- \\ \cline{1-1}
\multicolumn{1}{|c|}{$f_{\rm He}$} & \multicolumn{1}{c|}{---} & \multicolumn{1}{c|}{The re-scaling factor of He abundance} & \multicolumn{1}{c|}{(0.02, 5)} & --- \\ \cline{1-1}
\multicolumn{1}{|c|}{$f_{\rm C}$} & \multicolumn{1}{c|}{---} & \multicolumn{1}{c|}{The re-scaling factor of C abundance} & \multicolumn{1}{c|}{(0.2, 2)} & --- \\ \cline{1-1}
\multicolumn{1}{|c|}{$f_{\rm O}$} & \multicolumn{1}{c|}{---} & \multicolumn{1}{c|}{The re-scaling factor of O abundance} & \multicolumn{1}{c|}{(0.2, 2)} & --- \\ \hline
\end{tabular}%
}
\caption{The free and nuisance parameters for the propagation equations. All parameters are randomly scanned with a uniform distribution within corresponding ranges shown in the $4^{\rm th}$ column. The best-fit values in the $5^{\rm th}$ column are obtained by fitting the proton, helium, carbon, oxygen, boron-to-carbon and boron-to-oxygen ratios measured by ACE, AMS-02, DAMPE, and Voyager. 
}
\label{tab:Prior}
\end{table}

The \texttt{GALPROP}~\cite{1998ApJ...509..212S} program\footnote{\url{https://galprop.stanford.edu}} solves propagation equations in the interstellar medium (ISM) and calculates the LIS of CRs before they enter the solar system. Different CR particles are described by distinct but coupled propagation equations.

To consider the theoretical uncertainties from CR propagation and source parameters, 
\texttt{GALPROP} code is embedded with an MCMC sampling code \texttt{emcee}~\citep{2013PASP..125..306F} to perform a global fitting similar to Ref.~\cite{2023FrPhy..1844301M,2023RAA....23h5019X}. 
For the likelihoods, we include 
the spectrum data of proton, helium, boron-to-carbon and boron-to-oxygen ratios measured by AMS-02~\cite{2021PhR...894....1A} and DAMPE~\cite{DAMPE2019p, DAMPE2021He, DAMPE2022BC},
while the carbon and oxygen spectra data are taken from AMS-02~\cite{2021PhR...894....1A}.
The Voyager data outside the solar system~\cite{2016ApJ...831...18C, 2019NatAs...3.1013S} are also included.
The ACE-CRIS measurements\footnote{\url{ https://izw1.caltech.edu/ACE/ASC/level2/lvl2DATA_SIS.html}} including carbon, oxygen, boron-to-carbon and boron-to-oxygen ratios are used for the same period as AMS-02.
To reduce the impact of the degeneracy between the diffusion coefficient and the halo height, the ratios of $^{10}$Be/$^9$Be from several experiments are also included~\cite{2002ApJ...568..210W, 2001ApJ...563..768Y, 1988SSRv...46..205S, 1998ApJ...501L..59C, 2004ApJ...611..892H}.

With the likelihoods excluding the AMS-02 antiproton data, we scan 38 parameters, including 7 propagation parameters, 30 source parameters, and the solar modulation potential of CR. 
The prior ranges for the scanned parameters are shown in Table~\ref{tab:Prior}. 
For solar modulation, we use the FFA model, as detailed in Sec.~\ref{sec:FFA}. 
As a reference of correlations od the parapagation parameters, we provide the two-dimensional posterior contours of parameters in Fig.~\ref{fig:contour of propogation} of App.~\ref{sec:PPC}. 
The best-fit parameters are listed in the last column of Table~\ref{tab:Prior}, 
while their corresponding spectra are presented in Fig.~\ref{fig:galprop_results} of App.~\ref{sec:PPC}.

\subsection{The local interstellar spectrum of the antiproton}
\label{sec:LIS}

CRs interacting with atoms in ISM through high-energy collisions generate secondary particles including antiprotons and $\gamma$-rays. This process applies to both cosmic-ray protons ($H$) and heavier nuclei ($N$), resulting in numerous secondary particles including daughter nuclei with atomic numbers lower than those of the initial CRs or ISM gas particles.
Within the \texttt{GALPROP} framework, the following nuclear interaction channels are considered for secondary particle production calculations: $H$-$H$, $H$-$N$, $N$-$H$, and $N$-$N$. The corresponding cross sections for these processes are adopted from~\cite{1983JPhG....9..227T}.
The production and propagation mechanisms of antiprotons are implemented following the prescriptions detailed in~\cite{2002ApJ...565..280M} and~\cite{2015ApJ...803...54K}.
Using the optimized cosmic-ray spectra and propagation parameters from the previous section,
we calculate the LIS of secondary antiprotons produced via interactions of CRs with interstellar medium atoms.  
As a benchmark, the best-fit propagation parameter from Table~\ref{tab:Prior} yields the LIS of secondary antiprotons (red dashed line) shown in Fig.~\ref{fig:LIS vs. TOA}.

An additional source of antiprotons arises from DM annihilation processes. 
We model the DM halo distribution according to the Navarro-Frenk-White (NFW) profile~\cite{1997ApJ...490..493N}, $\rho(r) = \rho_s[(r/r_s)(1+r/r_s)^2]^{-1}$, with $r_s=20~\rm{kpc}$ and $\rho_s=0.35~\rm{GeV}~\rm{cm}^{-3}$, centered on the Galactic center, assuming DM annihilation through the $b\bar{b}$ channel.
The resulting antiproton injection spectrum is adopted from~\cite{2011JCAP...03..051C, Ciafaloni2011Mar}. 
Taking the best-fit propagation parameter from Table~\ref{tab:Prior}, 
we show the corresponding LIS of DM annihilation (the black dashed line) in Fig.~\ref{fig:LIS vs. TOA}.

\section{Cosmic ray propagation within solar system}
\label{sec:modulation}

When CRs enter the solar system, their spectra are modulated by the solar wind and heliospheric magnetic field, 
especially for those with rigidity below 40 GV. 
The 27.27-day solar rotation period (one BR cycle) produces corresponding periodic variations in CR fluxes, enabling their description between different BR cycles through solar modulation modeling.
The Parker transport equation~\cite{1965P&SS...13....9P} is commonly used to model the propagation of charged CRs in the solar system, 
accounting for interactions with the solar wind and heliospheric magnetic field via processes such as convection, drift, diffusion, and energy loss.
The equation is written as 
\begin{equation}
    \frac{\partial f}{\partial t}=
    -(\boldsymbol{V}_{\rm sw}+\boldsymbol{V}_{\rm d})\cdot\nabla f+
    \nabla \cdot(\boldsymbol{K}_{\rm s}\cdot\nabla f)+
    \frac{1}{3}(\nabla\cdot \boldsymbol{V}_{\rm sw} )\frac{\partial f}{\partial \ln{p}},
    \label{equation: ParkerEquation}
\end{equation}
where $f(\vec{r},p,t)$ is the phase space density distribution of CRs as a function of position $\vec{r}$, momentum $p$, and time $t$. 
The speed of the solar wind, the pitch-angle-averaged drift velocity, and the diffusion tensor are represented by $\boldsymbol{V}_{\rm sw}$, $\boldsymbol{V}_{\rm d}$, and $\boldsymbol{K}_{\rm s}$, respectively. Convection describes the transport of CRs by the solar wind, drift refers to the movement of particles due to the magnetic field gradient and curvature, diffusion is the random scattering of particles, and energy loss includes the various ways CRs lose energy during their journey.

The Parker transport equation can be numerically solved for this complex system, while it can be approximately solved with certain assumptions. In this work, we consider three models: (i) the FFA model~\cite{1968ApJ...154.1011G}, (ii) the extended FFA model~\cite{2016PhRvD..93d3016C, 2022JCAP...10..051C}, and (iii) numerical 3D simulation approach~\cite{2023ApJ...953..101A}, to solve solar modulation.
FFA is the simplest but widely used model that assumes a constant modulation potential, while the extended FFA model introduces more sophisticated treatments of the modulation potential, including time-, charge-, and rigidity-dependent modulation due to drifts in the heliospheric magnetic field. 
The extended FFA model, hereafter referred to as CT-dependent FFA model, improves accuracy at the cost of increased complexity in free parameters. 
Numerical 3D simulations, on the other hand, offer the most comprehensive and accurate approach by directly solving the Parker transport equation, despite a higher computational cost.

\subsection{Force-field approximation with a constant force-field energy loss}
\label{sec:FFA}

The FFA model~\cite{1968ApJ...154.1011G} provides an important analytical tool to calculate the top of the atmosphere (TOA) using CR intensity measured at the LIS,
\begin{equation}
    J^{\rm TOA}(E) = J^{\rm LIS}(E + \Phi) \times \frac{E (E + 2 m)}{(E + \Phi)(E + \Phi + 2m)}
    \label{equation:FFA}
\end{equation}
where $J^{\rm TOA}$ and $J^{\rm LIS}$ are the TOA and LIS of CRs, respectively. 
Here, $E$ is the particle kinetic energy per nucleon, $m$ is mass of proton, and $\Phi$ the force-field energy loss, given by $\Phi = Ze\phi / A$. 
In this expression, $Z$ and $A$ are the particle charge and mass number, $e$ is the elementary charge, and $\phi$ is the modulation potential, which typically ranges from 0.1 to 1 GV. For a detailed derivation of the FFA, see~\cite{1973Ap&SS..25..387G}.

\subsection{A time, charge, and rigidity-dependent force-field approximation}
\label{sec:model2}

\begin{figure}[!h]
    \centering
    \includegraphics[width=0.8\textwidth]{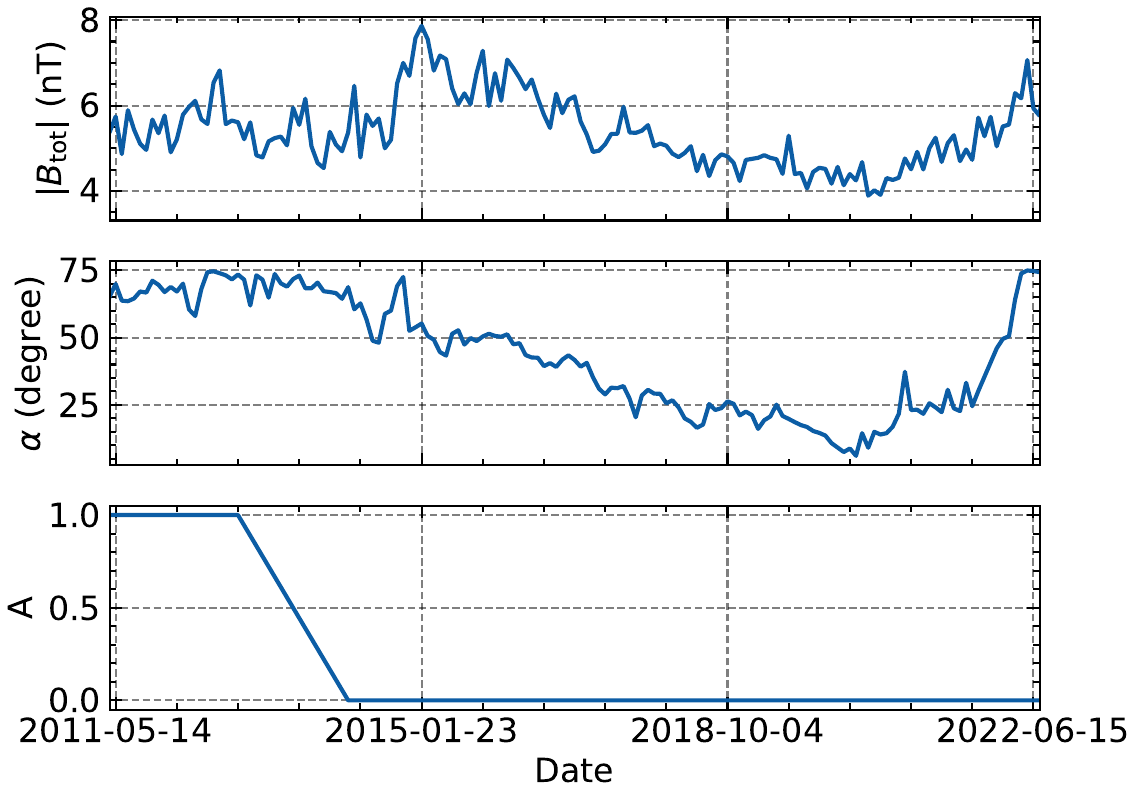}
    \caption{The time dependence of HMF $|B_{\rm tot}|$ (upper panel), HCS tilt angle $\alpha$ (middle panel), and polarity $A$ (bottom panel). The HMF $|B_{\rm tot}(t)|$ and polarity $A(t)$ are adopted from \url{https://izw1.caltech.edu/ACE/ASC/level2/lvl2DATA_MAG.html}, while $\alpha(t)$ is adopted from \url{http://wso.stanford.edu/Tilts.html}.
    The time span covers the period from May 2011 to June 2022.}
    \label{fig:HMF_Tilt_Polarity}
\end{figure}

The simple FFA does not account for charge-dependent effects from the heliospheric magnetic field, solar activity variability, and rigidity dependence of CRs. 
To address these shortcoming, Ref.~\cite{2016PhRvD..93d3016C} proposed an advanced modulation potential as a more comprehensive model that includes rigidity, time, and particle charge effects. 
Their analytical expression for $\phi$ using measurement data from multiple CR detectors is 
\begin{equation}
    \phi(R,t)=\phi_0\left( \frac{|B_{\rm tot}(t)|}{4 nT} \right)+
    \phi_1 H\left[x=-qA(t)\right] 
    \left( \frac{|B_{\rm tot}(t)|}{4nT} \right)
    \left[ \frac{1+\left( \frac{R}{R_0} \right)^2}{\beta\left( \frac{R}{R_0} \right)^3} \right]
    \left( \frac{\alpha(t)}{\pi/2} \right)^4, 
\end{equation}
where $\phi_0$ and $\phi_1$ are the time-independent parameters, the time-dependent quantities $|B_{\rm tot}(t)|$ and $A(t)$ represent the strength and polarity of the heliospheric magnetic field (HMF), 
while $\alpha(t)$ is the tilt angle of the heliospheric current sheet (HCS).   
The denominator of $|B_{\rm tot}(t)|$ is 4 nano-Tesla ($nT$).
As illustrated in Fig.~\ref{fig:HMF_Tilt_Polarity}, $|B_{\rm tot}(t)|$ (upper panel), $\alpha(t)$ (middle panel), and $A(t)$ (bottom panel) are functions of BR cycles. 
Additionally, $R$, $\beta$, and $q$ denote rigidity, velocity, and charge, respectively, with $R_0 = 0.5 \, \text{GV}$ as a reference rigidity. 
The \texttt{Heaviside step function} $H[x]$ accounts for the polarity inversion of the HMF, affecting particle propagation based on charge.

The modulation is implemented in a publicly available code~\footnote{\url{https://bitbucket.org/StockholmAstroparticle/solar-modulation/}}, enabling accurate predictions of solar modulation effects on charged CRs over time~\cite{2022JCAP...10..051C}.

\subsection{Three-dimensional numerical simulation of the Parker transport equation}
\label{sec:SDEs}

To numerically evaluate a comprehensive modulation potential, the Parker equation~\ref{equation: ParkerEquation} can be expressed as a set of stochastic differential equations (SDEs). 
For a pseudo-particle with rigidity $R$ at position $r$, this equivalent SDE formulation is given by 
\begin{equation}
    \begin{aligned}
        d \boldsymbol{r} &= \left(\nabla \cdot \boldsymbol{K}_{\rm s} - \boldsymbol{V}_{\rm sw} - \boldsymbol{V}_d\right) ds + \sqrt{2 \boldsymbol{K}_s} \cdot d \boldsymbol{W}(ds), \\
        d R &= \frac{1}{3} R \left(\nabla \cdot \boldsymbol{V}_{\rm sw}\right) ds,
    \end{aligned}
\end{equation}
where $ds$ represents backward time with $ds = -dt$. 
Using the Wiener process, each element in the random matrix $d \boldsymbol{W}$ follows a Gaussian distribution with zero mean and variance $ds$. 
As in Eq.~\eqref{equation: ParkerEquation}, the bold symbols $\boldsymbol{K}_{\rm s},~\boldsymbol{V}_{\rm sw},~\boldsymbol{V}_{\rm d}$ are the diffusion tensor, the speed of the solar wind, and the pitch-angle-averaged drift velocity, respectively.

The parallel and perpendicular diffusion of galactic CRs is the main contribution to small-scale irregularities in the HMF. 
For a simplicity, we adopt an empirical expression of $\boldsymbol{K}_{\rm s}$~\cite{2014SoPh..289..391P, 2020Ap&SS.365..182N, 2021ApJ...909..215A} as a $3\times 3$ matrix with diagonal terms, $\texttt{diag}(\boldsymbol{K}_{\rm s})=(K_{\|}, K_{\perp, r}, K_{\perp, \theta})$,    
\begin{equation}
    \begin{aligned}
         K_{\|} \equiv &~K_0 \beta k_1(r) k_2(R), \\
         K_{\perp, r} \equiv &~0.02 K_{\|},&& \\
         K_{\perp, \theta} \equiv &~\left ( 2 + \tanh\left[8 \left|\theta - 90^{\circ}\right| - 280^{\circ}\right]\right )\times K_{\perp, r},
    \end{aligned}
    \label{eq:K-matrix}
\end{equation}
where $K_0$ is a constant with units $10^{20} \, \text{cm}^2 \text{s}^{-1} $, while the $k_1$ and $k_2$ are dimensionless, as defined by    
\begin{equation}
    \begin{aligned}
        k_1(r) &\equiv \frac{B_{\mathrm{eq}}}{B}, \\
        k_2(R) &\equiv \left(\frac{R}{R_{k}}\right)^{a}
        \left[1+\left(\frac{R}{R_{k}}\right)^{\frac{b-a}{c}}\right]^{c},
    \end{aligned}      
    \label{eq:K-parallel}
\end{equation}
following Refs \cite{2014SoPh..289..391P, 1999ApJ...520..204G}.
The rigidity dependence slope is given by $a$ for rigidity below the turnover point $R_k$ and $b$ for rigidity above it. 
The scale parameter $c$ is fixed at 2.2 in this work, while the HMF strength near Earth is denoted by $B_{\text{eq}}$.

The HMF is an Archimedean spiral~\cite{1958ApJ...128..664P}, resulting from the Sun rotation. 
Still, the standard Parker HMF can cause excessive galactic CRs drift effects in the polar regions of heliosphere. 
Hence, a simple modification~\cite{1989GeoRL..16....1J}, to maintain $\nabla \cdot \boldsymbol{B} = 0$, 
introduces a small latitudinal component to the standard Parker HMF, 
\begin{equation}
    \begin{aligned}
        \boldsymbol{B} & =\frac{A_s B_0}{r^2}\left(\boldsymbol{e}_r+\xi \boldsymbol{e}_\theta-\Psi \boldsymbol{e}_{\varphi}\right) \times\left[1-2 H\left(\theta-\theta^{\prime}\right)\right], \\
        \xi & =\frac{r \delta_m}{r_s \sin \theta}, \\
        \Psi & =\frac{\left(r-r_s\right) \sin \theta \Omega}{V_{s w}}, 
    \end{aligned}
\end{equation}
where $H\left(\theta-\theta^{\prime}\right)$ again represents the \texttt{Heaviside function} with 
the HCS latitudinal extent introduced by Ref.~\cite{1981ApJ...243.1115J, 1983ApJ...265..573K} as 
\begin{equation}
    \theta^{\prime}=\frac{\pi}{2}-
    \arctan \left[\tan \alpha \sin \left(\varphi+\frac{\Omega \times \left(r-r_s\right)}{V_{\rm sw}}\right)\right].
\end{equation}
The parameter $B_0$ represents the HMF near Earth. 
The polarity of the HMF $A_s$ can be positive or negative, indicating whether the HMF points outward or inward in the Sun northern hemisphere. 
The Sun rotation speed is $\Omega = 2.66 \times 10^{-6}$ rad/s, and $\delta_m$ is set to $2 \times 10^{-5}$~\cite{2018AdSpR..62.2859B}. 
The perturbation parameter $\delta_m / \sin \theta$ ensures a divergence-free magnetic field.

For the solar wind velocity $V_{\rm sw}$, we summarize the findings of observations: 
\begin{itemize}
    \item The value of $V_{\rm sw}$ increases radially from zero to a constant value within 0.3 AU of the Sun~\cite{1997ApJ...484..472S}.
    \item During the minimum solar period, $V_{\rm sw}$ rises from approximately 400 km/s in the equatorial plane to 
    about 800 km/s at high heliolatitudes~\cite{2002GeoRL..29.1290M}.
    \item No significant latitude dependence is observed during the maximum solar period~\cite{2006SSRv..127..117H}.   
\end{itemize}
Therefore, the solar wind velocity throughout the heliosphere can be expressed as~\cite{2006SSRv..127..117H, 2013LRSP...10....3P, 2022PhRvD.106l3004S} 
\begin{equation}
    \boldsymbol{V}_{\rm sw} = V_0\left\{1-\exp \left[\frac{40}{3}\left(\frac{r_s-r}{r_0}\right)\right]\right\} \times\left\{1.475 \mp 0.4 \tanh \left[6.8\left(\theta-\frac{\pi}{2} \pm \xi\right)\right]\right\} \boldsymbol{e}_r,
\end{equation}
where $V_0$ represents the solar wind speed near Earth, $r_s = 0.005$ AU is the solar radius, $r_0 = 1$ AU, $\xi = \alpha + 15\pi/180$, and $\alpha$ denotes the tilt angle of the HCS. 
The upper and lower signs correspond to the northern and southern hemispheres, respectively.

\begin{figure}
    \centering
    \includegraphics[width=0.8\textwidth]{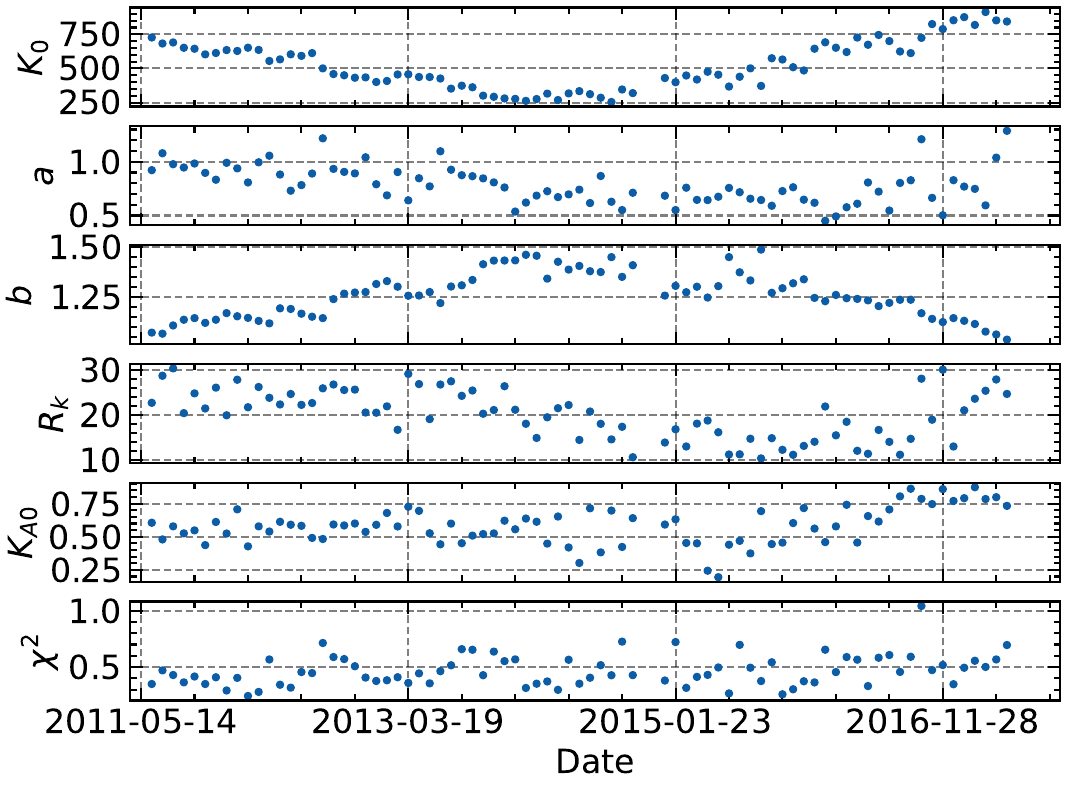}
    \caption{The time dependence of different parameters, 
    with $K_0$ in Eq.~\ref{eq:K-matrix}, $\{a, b, R_k\}$ in Eq.~\ref{eq:K-parallel}, and $K_{A0}$ in Eq.~\ref{eq:Vd}, and the reduced $\chi^2$ for 79 BR cycles proton flux measured by AMS-02~\cite{2018PhRvL.121e1101A}.
    }
    \label{fig:HMF_Tilt_SW}
\end{figure}

The drift velocity of charged particles arises from curvatures, gradients, and the current sheet in the large-scale HMF~\cite{2014SoPh..289..391P}, 
\begin{equation}
    \boldsymbol{V}_d=K_{A 0} \frac{q R \beta}{3} \nabla \times\left(\frac{\boldsymbol{B}}{B^2}\right),
    \label{eq:Vd}
\end{equation}
where $q$ is the particle charge-sign and $K_{A0}$ is a constant with a value between $0<K_{A0}<1$. 
If $K_{A0} = 1$, full drift case is described.

We extract the free parameters of simulation,   
\begin{equation}
    K_0,~a,~b,~R_{k},~{\rm and}~K_{A 0},  
    \label{eq:SunPS}
\end{equation}
from Eq.~\eqref{eq:K-matrix}, Eq.~\eqref{eq:K-parallel}, and Eq.~\eqref{eq:Vd} for the fitting process. 
These parameters are determined using monthly proton flux data from May 2011 to May 2017 measured by AMS-02~\cite{2018PhRvL.121e1101A}, and their best-fit values are shown in Fig.~\ref{fig:HMF_Tilt_SW}.
It's shown that $K_0$ and $b$ remain relatively stable, oscillating over a single period, 
while the other parameters fluctuate continuously. 
For a complementary, we present cosmic-ray proton fluxes with nine rigidity ranges as a function of time in App.~\ref{app:solar2017}.

We conducted extensive numerical simulations to determine the values of these parameters across multiple BR cycles. 
For the proton energy spectrum, we selected 45 discrete rigidity points per BR cycle, 
with 2,500 pseudo-particles assigned to each rigidity points in the simulation. 
To optimize the values of these five parameters, we used the MCMC sampling, generating five chains and running for 1,000 iterations. 
This procedure required approximately $4.4\times 10^{10}$ computational operations in total.

Following the determination of the parameter values for all BR cycles, we applied them to the solar modulation of antiprotons. 
Similar to the proton treatment, we selected 50 discrete rigidity points and reduced the number of pseudo-particles to 500 per rigidity point. 
The final computation involved propagating 10,000 distinct antiproton LIS through the solar modulation framework, generating corresponding TOA spectra.
This comprehensive analysis required approximately $2\times 10^{10}$ computational operations.

To scale the simulation results from the period before May 2017~\cite{2018PhRvL.121e1101A} to the period before June 2022~\cite{2025PhRvL.134e1002A}, 
we first note that our simulation covers only 79 BR cycles but needs to be extended to align with the AMS-02 dataset before June 2022. 
Due to the high computational cost associated with running additional simulations, 
we aim to predict the antiproton flux based on existing simulation results from the May 2017 observation period.
The process involves several key steps. First, we establish a rigidity-dependent scaling ratio, $r_i(R_i)$, by utilizing the CT-dependant FFA. 
This ratio allows us to compute the background fluxes for both the May 2017 and June 2022 periods using the AMS-02 best-fit modulation parameters specific to June 2022. 
Next, we compute the correction factors for each rigidity bin, where the ratio $r_i(R_i)$ is determined by comparing the averaged $\bar{p}$ flux for both periods. 
Finally, we scale the May 2017 baseline flux to obtain the predicted flux for June 2022 using the established correction factors, 
enabling an equivalent comparison with the AMS-02 data~\cite{2025PhRvL.134e1002A}.

\section{Results}
\label{sec:results}

According to the \texttt{emcee} code guidelines~\cite{2013PASP..125..306F}, the recommended convergence criterion is to run the chain for $50\tau$ steps, 
where $\tau$ is estimated from the autocorrelation function. 
This $\tau$ value indicates how many steps are needed for the chain to "forget" its initial state, ensuring effectively independent samples in MCMC.
In our scans, we collect approximately $10^6$ samples by reaching $50\tau$ steps. 
To minimize computation time, we then continue the scan to obtain a reduced data set of $10^5$ samples after convergence.
We utilize these $10^5$ samples for the DM signal and upper limit estimation.

\subsection{Modulation and Chi-square map}
\label{sec:chisq_map}

\begin{figure}[htbp]
    \centering
    \includegraphics[width=\textwidth]{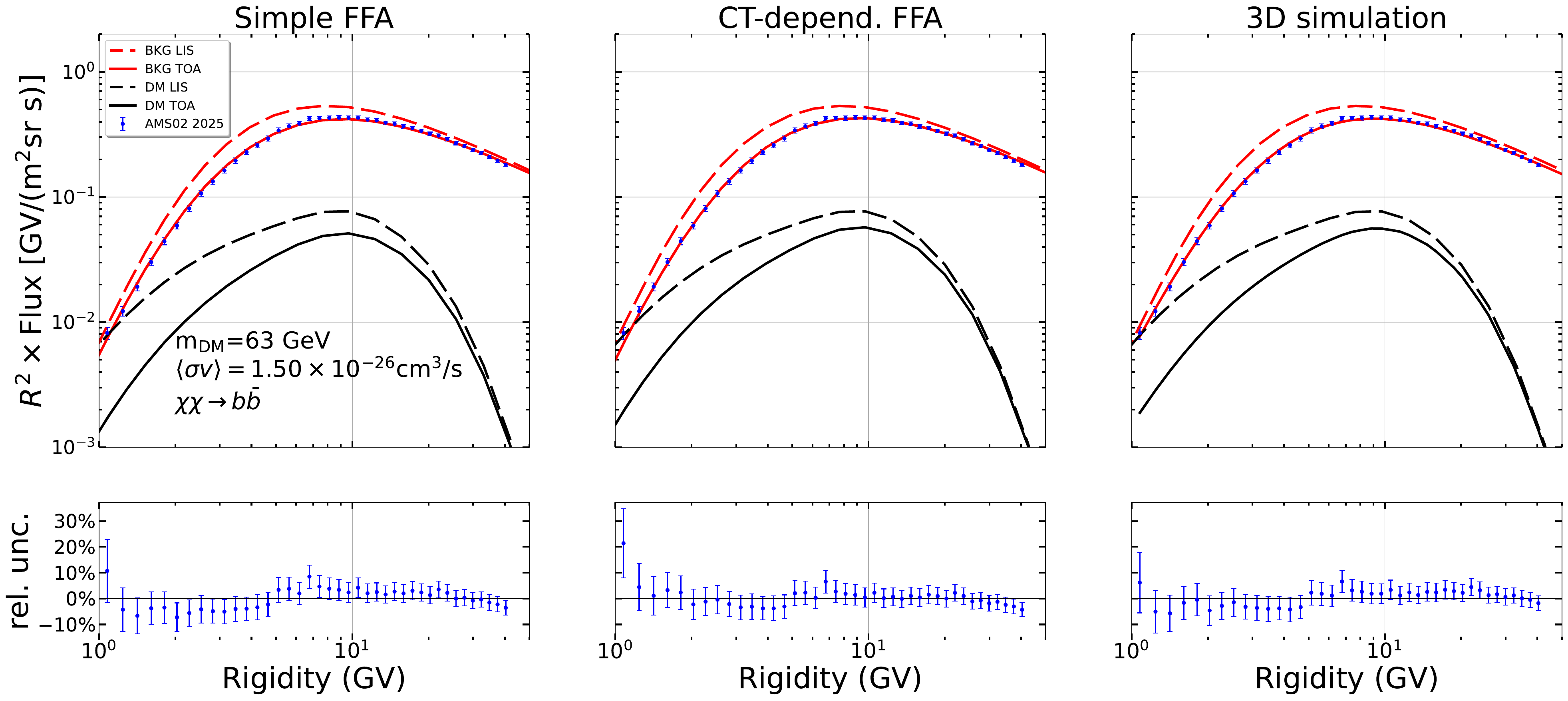}
    \caption{Comparison of antiproton flux predictions from three solar modulation models with AMS-02 data. 
    The three columns, from left to right, correspond to the FFA model, the CT-dependent FFA model, and the 3D numerical simulation approach. 
    In each panel, the red lines indicate the background flux, with solid lines representing the TOA flux and dashed lines representing the LIS flux, calculated using the best-fit propagation parameters from Table~\ref{tab:Prior}. 
The blue lines represent the additional flux from DM annihilation with a mass of 63~GeV, $b\bar{b}$ final states, and cross section of $1.5 \times 10^{-26}~\mathrm{cm^3/s}$.
    The lower subpanels display the residuals between the AMS-02 measurements and the background-only model.}
    \label{fig:LIS vs. TOA}
\end{figure}


To illustrate the impact of modulation, Fig.~\ref{fig:LIS vs. TOA} compares the LIS (dashed) and TOA (solid) antiproton spectra for the astrophysical background (red) and the DM contribution (blue). 
The spectra are computed with the best-fit propagation parameters from Table~\ref{tab:Prior} and shown for three solar modulation models: 
the FFA with $\Phi = 690$~MV (left), 
the CT-dependent FFA with $\phi_0 = 0.313$ and $\phi_1 = 4.8$ (middle), and 
the 3D numerical simulation (right) using parameters from Fig.~\ref{fig:HMF_Tilt_SW}. 
In all cases, the TOA spectra match the LIS spectra for $R > 40$ GV. 
The lower subpanels display the residuals between AMS-02 data and the background-only model, 
revealing a small excess near $7$~GV across all models, and an additional excess around $1$~GV in the CT-dependent FFA scenario.

\begin{figure}[htbp]
    \centering
    \includegraphics[width=\textwidth]{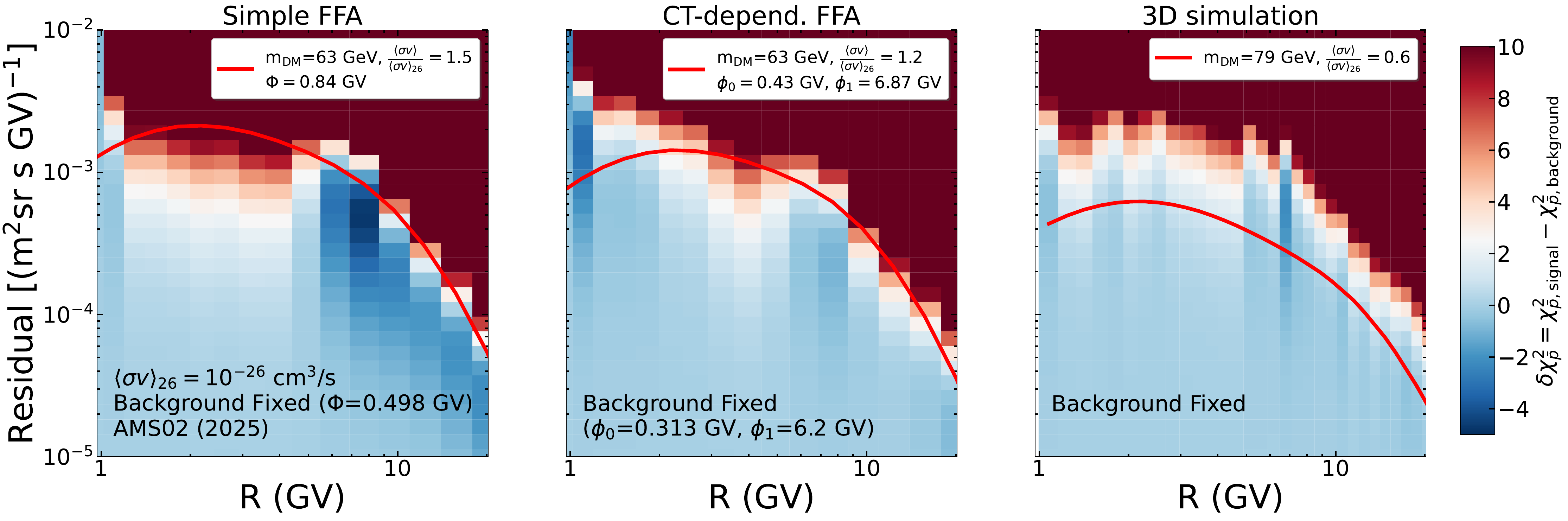}
    \caption{The $\delta\chi_{\bar{p}}^2$ map in the ($R$, residual flux) plane. 
    Based on the best TOA spectra of background antiprotons obtained using different solar modulation models fitted to the experimental data, three residual plots were generated. From left to right, the solar modulation models used are: FFA, CT-dependent FFA, and 3D simulation. Each plot includes a curve representing the TOA of $\bar{p}$ produced by DM.
    The color map illustrates the $\chi_{\bar{p}}^2$ difference between scenarios with and without a signal at each rigidity bin.}
    \label{fig:residual-map}
\end{figure}

In Fig.~\ref{fig:residual-map}, we project the $\delta\chi^2$ map onto the ($R$, residual flux) plane using the best-fit background and latest AMS-02 antiproton data.  
We define $\delta\chi_{\bar{p}}^2 = \chi^2_{\bar{p}, \text{signal}}-\chi^2_{\bar{p}, \text{background}}$
for each $R$ bin measured by AMS-02, 
where $\chi^2_{\bar{p}, \text{background}}$ is the minimum $\chi^2$ value fitted with the background, while $\chi^2_{\bar{p}, \text{signal}}$ is calculated with the background plus a residual flux from $10^{-5}$ to $10^{-2}~\text{m}^{-2}~\text{s}^{-1}~\text{sr}^{-1}~\text{GV}^{-1}$.
The $\delta\chi_{\bar{p}}^2$ values are presented in the color bar.

In the three panels, we also show three benchmark antiproton spectra corresponding to different DM masses and annihilation cross-sections: the left one for \{$m_{\rm DM} = 63$ GeV, $\sv = 1.5 \times 10^{-26}~\text{cm}^3~\text{s}^{-1}$\}, the middle one for \{$m_{\rm DM} = 63$ GeV, $\sv = 1.2 \times 10^{-26}~\text{cm}^3~\text{s}^{-1}$\}, and the right one for \{$m_{\rm DM} = 79$ GeV, $\sv = 6.0 \times 10^{-27}~\text{cm}^3~\text{s}^{-1}$\}.

In the left panel of Fig.~\ref{fig:residual-map}, increasing the residual to around $5 \times 10^{-4}~\text{m}^{-2}~\text{s}^{-1}~\text{sr}^{-1}~\text{GV}^{-1}$ near 8 GV significantly improves the fit to AMS-02 antiproton data, reducing the background $\chi^2$ by up to 4.7 units. 
This can be explained by a DM annihilation signal with $m_\chi = 63$ GeV and $\sv = 1.5 \times 10^{-26}~\text{cm}^3~\text{s}^{-1}$. 
In the middle and right panels of Fig.~\ref{fig:residual-map}, the signal regions shift to $R \sim 1.5$~GV for the CT-dependent FFA approach and $R \sim 7$~GeV for the 3D simulation approach. 

\subsection{Systematic uncertainties}
\label{sec:sys}

To address systematic uncertainties from AMS-02 data correlations, 
we adopt two approaches: (i) the add-in-quadrature method and (ii) the nuisance parameter method.

In the add-in-quadrature method, we assume the systematic uncertainties between different energy bins are uncorrelated.
The total uncertainty $\sigma_{\rm total}$ used in our antiproton likelihood is  
\begin{equation}
\sigma_{\rm total}^2 = \sigma_{\rm stat}^2 + \sigma_{\rm sys}^2. 
\end{equation}
The values of the statistical and systematic uncertainties ($\sigma_{\rm stat}$ and $\sigma_{\rm sys}$) for the entire $\bar{p}$ spectrum are provided in Ref.~\cite{2025PhRvL.134e1002A}.

The nuisance parameter method is more involved. 
Following Ref.~\citep{2017PhRvD..95h2007A}, we model systematic uncertainties as unknown correction factors, which are treated as nuisance parameters in the antiproton spectrum fit. 
The $\chi^2$ function of AMS-02 $\bar{p}$ data is 
\begin{equation}
\chi^2 = \sum_{i=1}^{n} \left(
\frac{F_i - [1+s(E_i;w)S_i] \Phi(E_i;\theta)}{\sigma_{{\rm stat},i} } \right)^2 + 
\sum_{j=1}^{N}w_j^2,
\end{equation}
where $n$ is the number of energy bins. 
For each bin $i$, $F_i$ and $\Phi$ represent the measured and predicted fluxes, 
$S_i$ is the relative systematic uncertainty, and $\sigma_{{\rm stat},i}$ contains only the statistical uncertainty. 
The function $s(E_i; w)$ is a piecewise-linear function of $\log_{10} E$, parameterized by its values $w_j$ at reference energies $\epsilon_j$. 
These $\epsilon_j$ are $N$ logarithmically spaced reference energies between 1~GV and 41.9~GV, with the $w_j$ values serving as nuisance parameters. 
The second term in the $\chi^2$ function represents a Gaussian prior on the amplitude of these nuisance parameters.
In this work, we scan the nuisance parameter number $n$ from 2 to 20 and find that the reduced $\chi^2$ is minimized at $n=7$.

Both methods are used in the antiproton flux modulation analysis for DM searches. 
The add-in-quadrature method results in larger uncertainties, leading to a lower reduced chi-square ($\chi^2/dof$) and 
a more conservative treatment of systematic uncertainties. 
In contrast, the nuisance parameter method only includes statistical error bars 
in the denominator when calculating $\chi^2/dof$.
In principle, 
including more nuisance parameters can decrease $\chi^2/dof$, although our best-fit values still show $\chi^2/dof > 1.5$. 
This suggests that the systematic uncertainties represented by the nuisance parameters are inaccurate, likely due to the misuse of uncorrelated uncertainties. 
As noted in Fig. 2 of Ref. \cite{2020PhRvR...2d3017H}, distinguishing between correlated and uncorrelated systematic uncertainties of AMS-02 is challenging. 
Consequently, the nuisance parameter method can generate more pessimistic results, producing a higher $\chi^2/dof$.

\subsection{Regions of DM signal}
\label{sec:excess_region}

\begin{figure}[ht!]
    \centering
    \includegraphics[width=\linewidth]{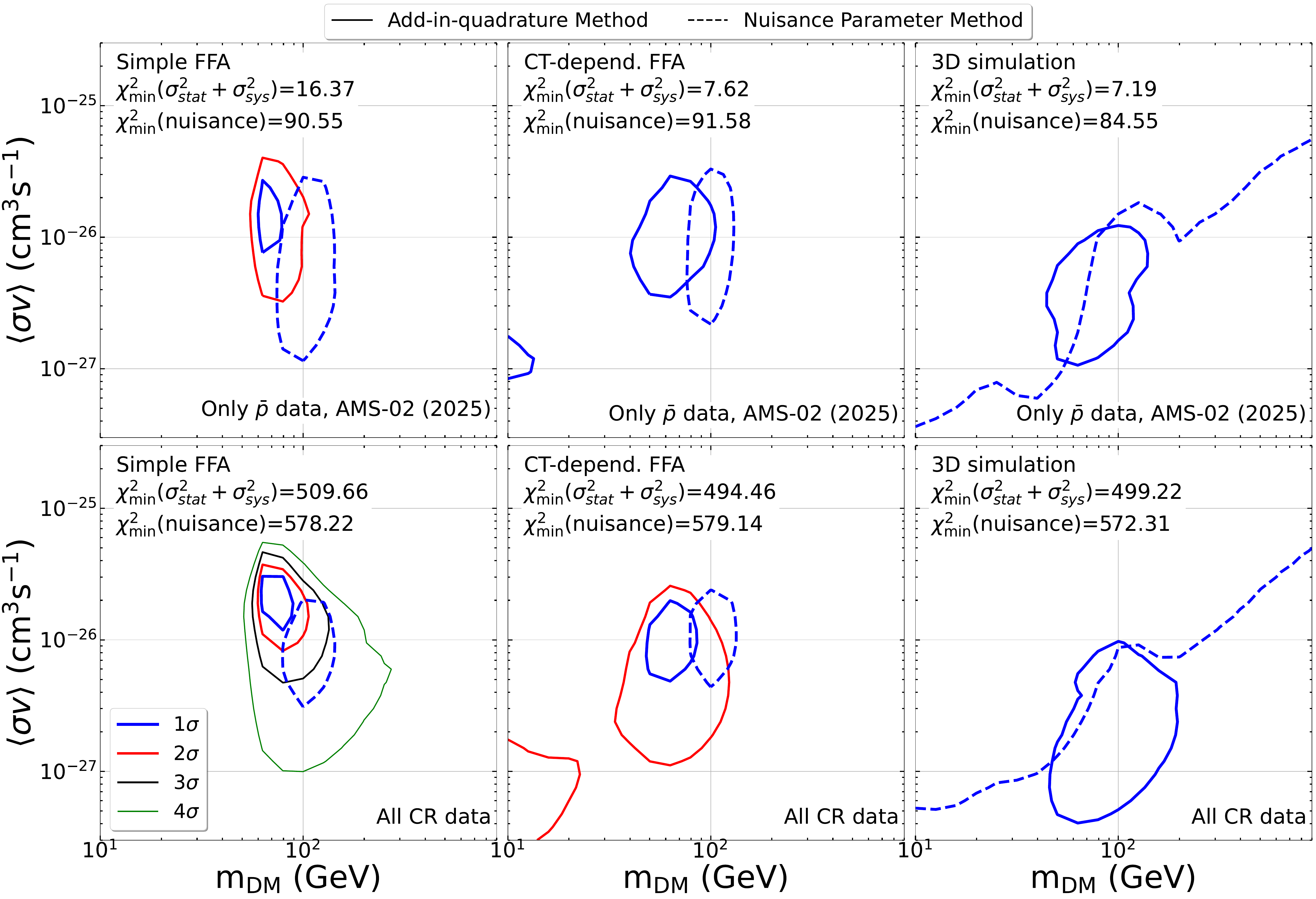}
    \caption{Contour plots of DM mass versus annihilation cross-section, 
    using the simple FFA model (left panels), the CT-dependent FFA model (middle panels), and the 3D simulation approach (right panels) 
    to account for solar modulation effects on CRs, respectively. 
    The upper three panels show that $\delta\chi^2$ accounts solely antiproton experimental data (only $\bar{p}$ data), 
    while the lower three panels combine data from antiprotons, protons, helium, and other CRs (all CR data). 
    The solid and dashed lines represent the add-in-quadrature and nuisance parameter approaches for addressing systematic errors, respectively.
    Different colored contours from inner to outer represent the 
    $1\sigma$ (\mkblue{blue}), 
    $2\sigma$ (\mkred{red}), 
    $3\sigma$ (black), and
    $4\sigma$ (\mkgreen{green}) confidence regions 
    for a $\chi^2$ distribution with 2 degree of freedom.}
    \label{fig:chi2_counter_2025}
\end{figure}

Unlike the procedure for determining CR parameters (see Tab.~\ref{tab:Prior}), 
searching for DM annihilation signals from antiproton data may rely on prior assumptions. 
Additionally, numerically solving the Parker transport equation is still CPU-intensive. 
To address these issues, we estimate statistical strength by post-processing the selected data set of $10^5$ samples using a two-dimensional grid scan on the ($m_{\rm DM}$, $\sv$) plane with the profile likelihood method. 
Considering Gaussian likelihoods for all data, we apply significance criteria in the two-dimensional parameter space: $\delta\chi^2 = 2.30$ for $1\sigma$, $\delta\chi^2 = 6.18$ for $2\sigma$, $\delta\chi^2 = 11.83$ for $3\sigma$, and $\delta\chi^2 = 19.33$ for $4\sigma$, where $\delta\chi^2 = \chi^2 - \chi^2_{\rm min}$.

Fig.~\ref{fig:chi2_counter_2025} shows the two-dimensional likelihood contour map projected onto the ($m_{\rm DM}$, $\sv$) plane, 
based on three solar modulation models: the simple FFA model described in Sec.~\ref{sec:FFA} (left two panels), 
the CT dependent FFA model described in Sec.~\ref{sec:model2} (middle two panels), 
and the 3D simulation approach detailed in Sec.~\ref{sec:SDEs} (right two panels).
The upper and lower panels display the statistical strength $\chi^2$ for the AMS-02 antiproton data and for the full data set, including $\bar{p}$ and other CRs DATA (H, He, C, O, B/C, B/O), respectively. 
In all panels, systematic errors are accounted for using the add-in-quadrature method (solid lines) and the nuisance parameter method (dashed lines). 
The four contours correspond to different confidence levels: 
    $1\sigma$ (\mkblue{blue}), 
    $2\sigma$ (\mkred{red}), 
    $3\sigma$ (black) and 
    $4\sigma$ (\mkgreen{green}). 
Only closed contours are shown, excluding those containing vanished $\sv$. 
The $\chi^2$ values for the best fits of two contours are provided in the legends.

Comparing the two methods for implementing systematic uncertainties in Fig.~\ref{fig:chi2_counter_2025}, 
the nuisance parameter method (dashed contours) results in less significant signals (up to 1$\sigma$). 
Except for the FFA model with all CR data (lower left panel), 
most signals found through the add-in-quadrature method are around 1-2$\sigma$. 
This is because the uncertainties of AMS-02 antiproton data are dominated by systematic errors, 
and adding them in quadrature leads to a conservative estimate. 
Note that the signal regions for the nuisance parameter method lies around 100 GeV, 
while for the add-in-quadrature method, it is mainly below 100 GeV. 
This discrepancy arises from the way that energy bins are patched in the nuisance parameter method, 
where forcing a correlation between bins can shift the signal region.

For solar modulations using simple FFA and CT-dependent FFA models, the signal strength increases when all CR data are included. 
This highlights that the antiproton signal strengthens once the propagation and source parameters are well constrained, 
regardless of how systematic uncertainties are implemented. 
However, for solar modulation using 3D simulation approach, the two ways for implementing systematic uncertainties behave differently. 
In the nuisance parameter method, closed contours cannot be formed for either dataset.
In contrast, the add-in-quadrature method keeps the signal at the $1\sigma$ level, while 
it is difficult to draw statistical conclusions even with all data included.

As shown in Fig.~\ref{fig:residual-map}, there is a small excess around $R \approx 2$~GV for the CT-dependent FFA. 
This suggests that DM annihilation with $m_{\rm DM} \sim 10$~GeV fits well in Fig.~\ref{fig:chi2_counter_2025}, 
though the signals have the lowest significance.

\subsection{Best-fits and 95~$\%$ upper limits}
\label{sec:BF}

\begin{table}[!h]
\centering
\begin{tabular}{|c|c|c|c|c|}
\hline
\multicolumn{5}{|c|}{Add-in-quadrature method} \\
\hline
\multicolumn{2}{|c|}{\shortstack{$\chi^2_{\rm min}/$dof\\(m$_{\rm DM}$/GeV, $\sv$/10$^{-26}~$cm$^3$~s$^{-1}$)}}  
& \multirow{-2}{*}{FFA} 
& \multirow{-2}{*}{CT-dep. FFA} 
& \multirow{-2}{*}{3D simulation} \\ 
\hline

\multirow{3}{*}{Only $\bar{p}$ data} 
& Bkg. 
& 27.46/39 = 0.70 
& 13.33/38 = 0.35
& 27.58/40 = 0.69\\
\cline{2-5}
& \multirow{-2}{*}{Bkg.+DM}
& \shortstack{\rule{0pt}{1.5em}16.37/37 = 0.44 \\ (63.1, 1.5)} 
& \shortstack{7.62/36 = 0.21 \\ (63.1, 0.95)}
& \shortstack{7.19/38 = 0.19 \\ (79.4, 0.60)} \\

\hline
\multirow{3}{*}{All CR data} &
Bkg. 
& 531.16/588 = 0.90 
& 501.52/587 = 0.85
& 519.43/589 = 0.88\\
\cline{2-5}
& \multirow{-2}{*}{Bkg.+DM}
& \shortstack{\rule{0pt}{1.5em}509.66/586 = 0.87 \\ (79.4, 1.9)}
& \shortstack{494.46/585 = 0.85 \\ (63.1, 1.2)}
& \shortstack{499.22/587 = 0.85 \\ (79.4, 0.24)} \\ 
\hline

\hline
\multicolumn{5}{|c|}{Nuisance parameter method} \\

\hline
\multicolumn{2}{|c|}{\shortstack{$\chi^2_{\rm min}/$dof\\(m$_{\rm DM}$/GeV, $\sv$/10$^{-26}~$cm$^3$~s$^{-1}$)}} 
& \multirow{-2}{*}{FFA} 
& \multirow{-2}{*}{CT-dep. FFA} 
& \multirow{-2}{*}{3D simulation} \\ 
\hline
\multirow{3}{*}{Only $\bar{p}$ data} 
& Bkg. 
& 93.91/32 = 2.93 
& 95.67/31 = 3.09 
& 86.13/33 = 2.61\\
\cline{2-5}
& \multirow{-2}{*}{Bkg.+DM}
& \shortstack{\rule{0pt}{1.5em}90.55/30 = 3.02 \\ (100.0, 0.75)}
& \shortstack{91.58/29 = 3.16 \\ (100.0, 1.2)}
& \shortstack{84.55/31 = 2.73 \\ (100.0, 0.48)} \\

\hline
\multirow{3}{*}{All CR data}
& Bkg. 
& 583.09/581 = 1.00
& 585.30/580 = 1.01
& 573.05/582 = 0.98\\
\cline{2-5}
& \multirow{-2}{*}{Bkg.+DM}
& \shortstack{\rule{0pt}{1.5em}578.22/579 = 1.00 \\ (100.0, 0.95)}
& \shortstack{579.14/578 = 1.00 \\ (100.0, 1.2)}
& \shortstack{572.31/580 = 0.99 \\ (100.0, 0.24)} \\ 
\hline
\end{tabular}
\caption{Best-fit values of reduced $\chi^2$ and DM parameters ($m_{\rm DM}$, $\sv$) shown in Fig.~\ref{fig:chi2_counter_2025}.}
\label{tab:Best-fit}
\end{table}

Tab.~\ref{tab:Best-fit} presents the best-fit reduced $\chi^2$ values for different solar modulation models, CR datasets, and systematic error treatment methods. 
The add-in-quadrature method generally underestimates the reduced $\chi^2$ values, 
while the nuisance parameter method yields values between 2.6 and 3.2, reflecting larger systematic errors than statistical ones. 
Regardless of the method, the total $\chi^2$ values with DM signal are always better than the background-only fits. 
Due to the large degree of freedom, the reduced $\chi^2$ values are within an acceptable range when all CR data are included.

As discussed in Sec.~\ref{sec:sys}, the nuisance parameter method can be pessimistic due to the misapplication of uncorrelated components within correlated systematic uncertainties. 
While including the DM signal slightly improves the total $\chi^2$ value, 
it adds two degrees of freedom, resulting in slight improvement in the reduced $\chi^2$ value, 
which remains outside the acceptable range of $0.5 < \chi_{\rm min}^2/{\rm dof} < 1.5$. 
This further suggests that we are misusing uncorrelated components within the correlated systematic uncertainties. 
In contrast, the add-in-quadrature method, which misapplies the systematic uncertainties as a mixture of correlated and uncorrelated components, produces well-fitting results, with the reduced $\chi^2$ falling within this range.

\begin{figure}[ht!]
    \centering
    \includegraphics[width=\textwidth]{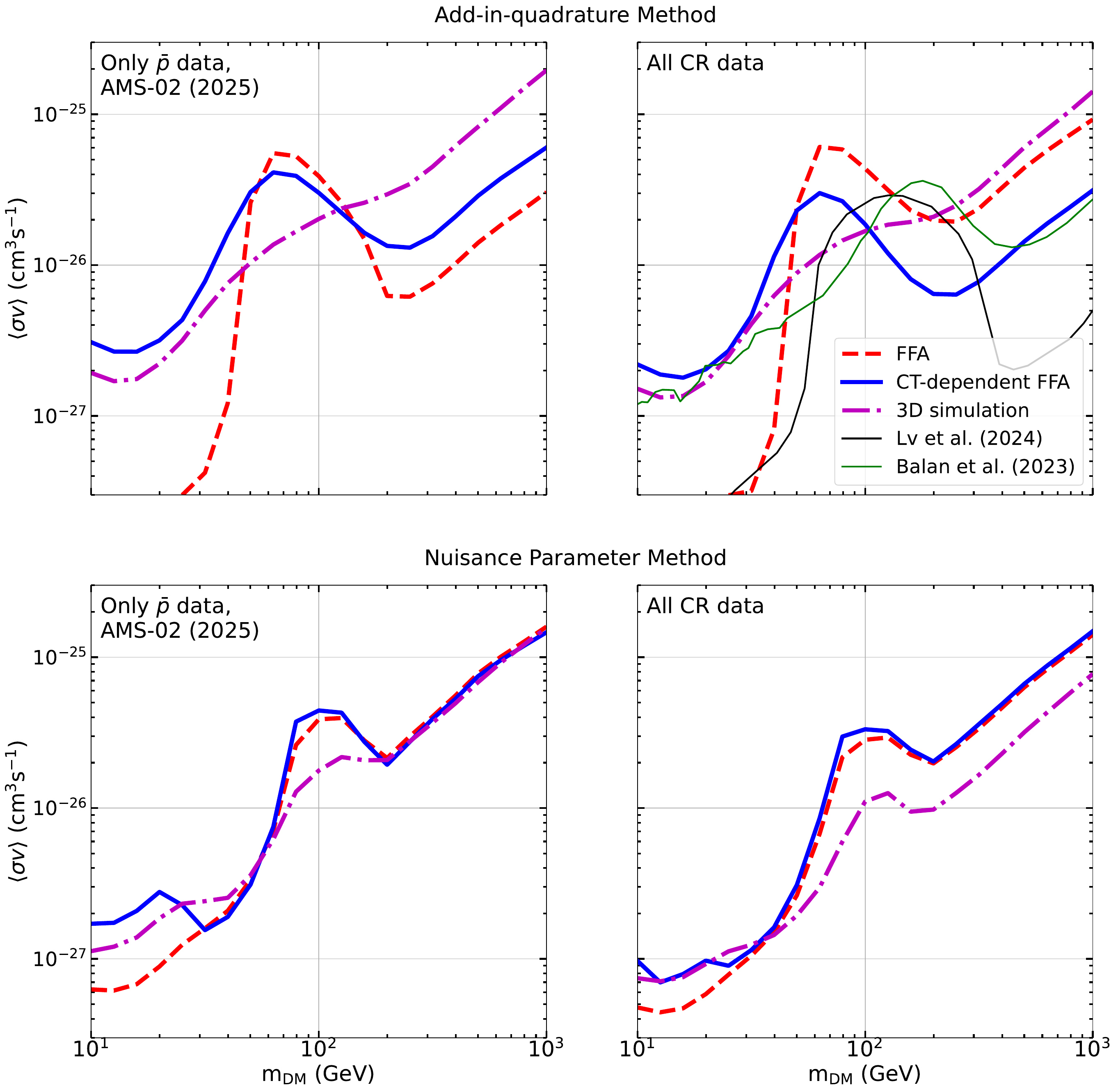}
    \caption{The 95\% upper limits of the DM annihilation cross section versus mass derived by fitting the AMS-02 data with different systematic error treatment methods, different modulation methods, and different experimental data. Different modulation methods are represented by different colored lines: simple FFA model (red dashed line), CT-dependent FFA model (blue solid line), and 3D simulation approach (purple dotted-dashed line). 
    The black and green solid lines in the upper right panel are taken from Fig.~5 of \cite{2024PhRvD.109d3006L} and Fig.~3 of \cite{2023JCAP...08..052B}, respectively.
    }
    \label{fig:upperlimits}
\end{figure}

In Fig.~\ref{fig:upperlimits}, we derive the 95\% upper limits of the DM annihilation cross-section for DM masses ranging from 10 to $10^3$ GeV with $\delta\chi^2=\chi^2_{\rm Bkg.+DM}-\chi^2_{\rm Bkg.}=2.71$, where $\chi^2_{\rm Bkg.+DM}$ and $\chi^2_{\rm Bkg.}$ are the $\chi^2$ values fitted with DM and without DM, respectively. 
The upper panels display these limits using three solar modulation models: the simple FFA model (red dashed line), the CT-dependent FFA model (blue solid line), and the 3D simulation approach (magenta dashed-dot line), with systematic uncertainties incorporated via the add-in-quadrature method. 
The lower panels show the corresponding limits obtained using the nuisance parameter method. 
The left panels present results from the antiproton-only analysis, while the right panels include all CR datasets.
For comparison, we also include results from previous studies: the constraints from Fig. 5 of Ref.~\cite{2024PhRvD.109d3006L} assuming a halo height of $L = 7.17~\rm{kpc}$ (black thin solid line), and those from Fig. 3 of Ref.~\cite{2023JCAP...08..052B} derived using their injection break model (green thin solid line).

We summarize our findings of Fig.~\ref{fig:upperlimits} as follows:
\begin{itemize}
    \item The add-in-quadrature method produces various upper limits for all three solar modulation models, 
    while the nuisance parameter method yields consistent results.
    
    \item The impact of systematic uncertainties on the upper limits is smaller than the impact of solar modulation effects.
    
    \item Compared with Fig.~\ref{fig:chi2_counter_2025}, a bump feature around 100~GeV correlates with signal significance. 
    A larger bump, as seen with the simple FFA model, corresponds to higher significance, 
    whereas a weaker signal like the 3D simulation approach results in a smaller bump.
    
    \item In the low-mass region, the upper limits derived from both the CT-dependent FFA model and the 3D simulation approach are higher than those from the simple FFA model. This difference arises because the CT-dependent FFA model and 3D simulation approach predict a non-zero signal in the low-energy regime (see Fig.~\ref{fig:residual-map}), while the simple FFA model shows no excess in this range. 
\end{itemize}

\section{Conclusion}
\label{sec:conclusion}

The latest AMS-02 measurements reveal a significant dependence of proton and antiproton fluxes on both charge and rigidity throughout the 11-year solar cycle. 
These observations offer valuable insights into the modulation mechanisms of charged particles by the solar wind and the heliospheric magnetic field. 
Motivated by this, we have comprehensively examined the impact of solar modulation on the antiproton flux excess, using three distinct solar modulation models: 
the simple FFA model, the time-, charge-, and rigidity-dependent FFA model, and the more detailed 3D numerical simulation approach based on the Parker transport equation. 
To identify potential signals of DM annihilation, we applied the profile likelihood method and considered systematic uncertainties using both the add-in-quadrature and nuisance parameter methods.

Our results highlight that the significance of the DM signal is highly sensitive to the solar modulation model chosen. 
For instance, using the add-in-quadrature method, a DM signal with a significance of approximately 2$\sigma$ is observed when applying the simple FFA model. However, this significance decreases to 1$\sigma$ when the CT-dependent FFA model or the 3D simulation approach are used.
The reduced $\chi^2$ values indicate that the add-in-quadrature method may wrongly estimate uncertainties due to its assumption of uncorrelated systematic errors. 
In contrast, the nuisance parameter method maintains a 1$\sigma$ significance across both FFA and CT-dependent FFA models, 
while no significant signal is found using the 3D simulation approach. 
The nuisance parameter method, though yielding larger reduced $\chi^2$ values, offers a more consistent estimation of DM parameters across different modulation models.

For three solar modulation models fitted to AMS-02 antiproton data, 
the add-in-quadrature method results in overfitting, while the nuisance parameter approach leads to underfitting. 
We found that the FFA model with the add-in-quadrature method is the statistically most reliable. 
When all CR data are included, the significance of the DM annihilation signal increases. 
With the add-in-quadrature method, the signal strength reaches $4\sigma$ in the simple FFA model, 
but drops to $2\sigma$ for the CT-dependent FFA and $1\sigma$ for the 3D simulation models. 
Using the nuisance parameter method, the significance remains at $1\sigma$ for both FFA and CT-dependent FFA models, while the 3D simulation approach shows no signal. 
The reduced $\chi^2$ values approach unity when including all CR data into the likelihood, indicating the dominance of the heavy CR data.

Despite the theoretical uncertainties related to CR propagation and antiproton production, 
the observed excess remains compatible with DM annihilation models, particularly those predicting a mass of approximately $63-79$ GeV and an annihilation cross-section of the order $\langle \sigma v \rangle \sim 10^{-26} \, \text{cm}^3\text{s}^{-1}$. 
While the add-in-quadrature method produces lower reduced $\chi^2$ values, the best-fit DM parameters differ based on the solar modulation model used. 
In contrast, the nuisance parameter method results in higher reduced $\chi^2$ values but offers more consistent estimates for the DM parameters among the three modulation models.

In addition to examining the signal region, we also investigated the effects of modulation models and data correlations on the derived upper limits of the DM annihilation cross-section. 
Our findings reveal that the 95\% upper limits from the simple FFA model present a significant bump around 100 GeV, indicating increased signal significance. Importantly, our results show that systematic uncertainties have a less impact compared to solar modulation effects on these upper limits. 
Moreover, the nuisance parameter method produced more consistent results among models than the limits derived from the add-in-quadrature approach. 
For DM mass around 100 GeV, $\langle \sigma v \rangle$ must be less than $3\times 10^{-26}\, \text{cm}^3\text{s}^{-1}$, regardless of the modulation models and usages of systematic uncertainties.

While this work highlights the potential for DM annihilation to explain the observed antiproton excess, 
the precise characterization of systematic uncertainties---especially those related to the solar modulation model and correlations in the data---remains crucial. 
Further refinement in modeling and a more accurate treatment of systematic uncertainties could provide a more definitive answer regarding the DM interpretation of the excess.

\section*{Acknowledgments}
This work is supported by the National Key Research and Development Program of China (No. 2022YFF0503304 and 2022YFF0503303), the Project for Young Scientists in Basic Research of the Chinese Academy of Sciences (No. YSBR-092), and the National Natural Science Foundation of China (No. 12073057). 
We sincerely thank Professor Qiang Yuan, Fei Gao, Zu-Hao Li and Cheng-Rui Zhu for discussion.


\appendix

\section{Supplementary figures of propagation parameters and fitted results of proton, helium, carbon, oxygen and secondary-to-primary spectra}
\label{app:ppc}

\begin{figure}[h]
    \centering
    \includegraphics[width=\textwidth]{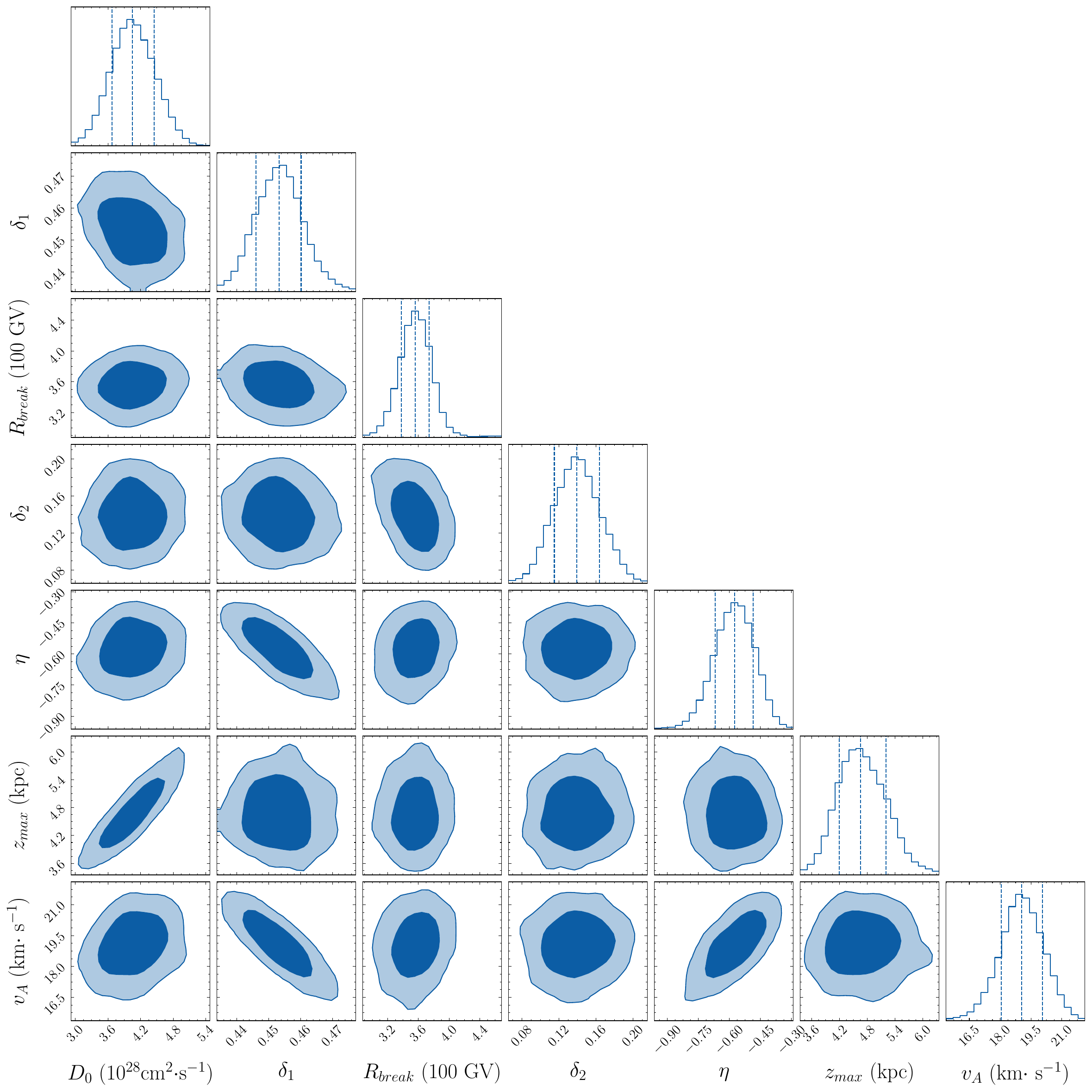}
    \caption{The distribution of the 100,000 sets of propagation parameters used in this study. There are a total of 7 parameters related to propagation, including diffusion coefficient, first diffusion index, broken rigidity, second diffusion index, velocity index, Galactic height, and Alfvén velocity. Using these parameters that have already stabilized as benchmarks, this work is completed.}
    \label{fig:contour of propogation}
\end{figure}

We perform a comprehensive spectral analysis of proton, helium, carbon, oxygen, and secondary-to-primary ratios measured by ACE, AMS-02, DAMPE, and Voyager.
These spectra are modeled using the GALPROP code. The MCMC analysis yielded the propagation parameter distributions shown in Fig.~\ref{fig:contour of propogation}.
There are 7 parameters related to propagation, including diffusion coefficient, first diffusion index, broken rigidity, second diffusion index, velocity index, Galactic height, and Alfvén velocity. 

\begin{figure}[!h]
    \centering
    \includegraphics[width=0.3\textwidth]{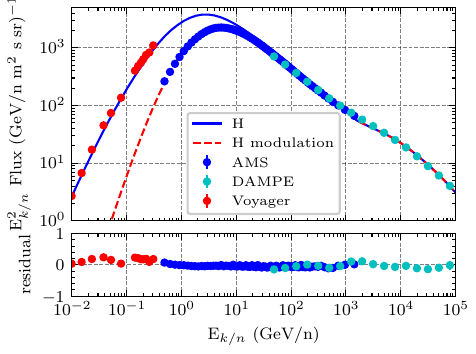}
    \includegraphics[width=0.3\textwidth]{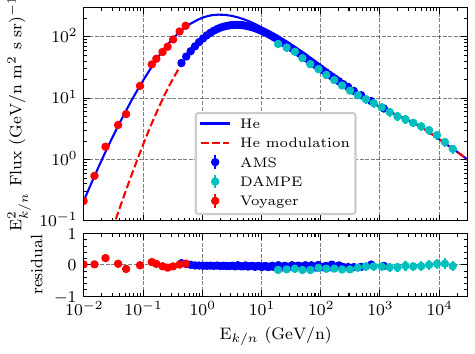}
    \includegraphics[width=0.3\textwidth]{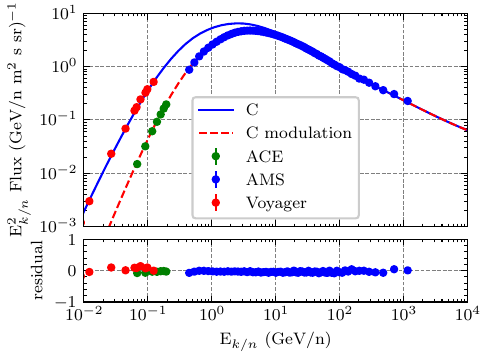}
    \includegraphics[width=0.3\textwidth]{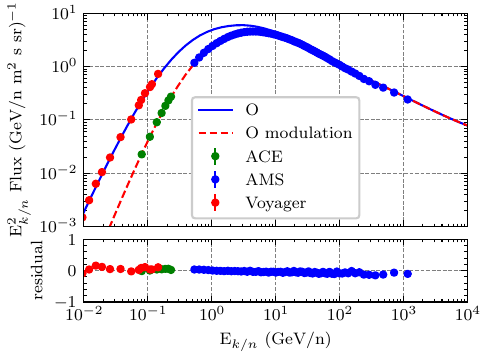}
    \includegraphics[width=0.3\textwidth]{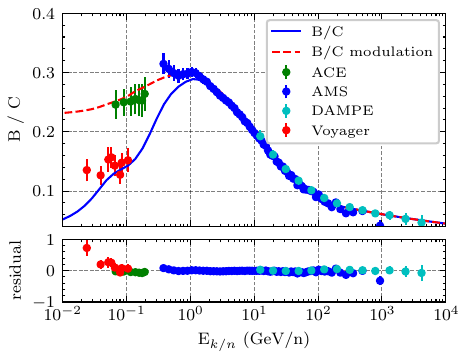}
    \includegraphics[width=0.3\textwidth]{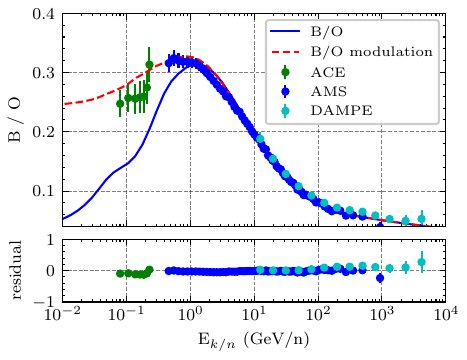}
    \caption{The fitted results of proton, helium, carbon, oxygen and secondary-to-primary spectra compared with data meansured by ACE, AMS-02, DAMPE and Voyager. In all panels, blue solid lines are LIS, while red dashed lines are TOA. Different points correspond to different experiments data: ACE (\mkgreen{green}), AMS-02 (\mkblue{blue}), DAMPE (\mkcyan{cyan}) and Voyager (\mkred{red}).}
    \label{fig:galprop_results}
\end{figure}

The best-fit spectra are shown in Fig.~\ref{fig:galprop_results}, demonstrating excellent agreement with the experimental data. 
In Fig.~\ref{fig:galprop_results}, blue solid lines are LIS, while red dashed lines are TOA. Different points correspond to different experiments data: ACE (\mkgreen{green}), AMS-02 (\mkblue{blue}), DAMPE (\mkcyan{cyan}) and Voyager (\mkred{red}).

\section{Goodness of Fit for AMS02 Monthly Data between 2011 and 2017}
\label{app:solar2017}
\begin{figure}[!h]
    \centering
    \includegraphics[width=0.3\textwidth]{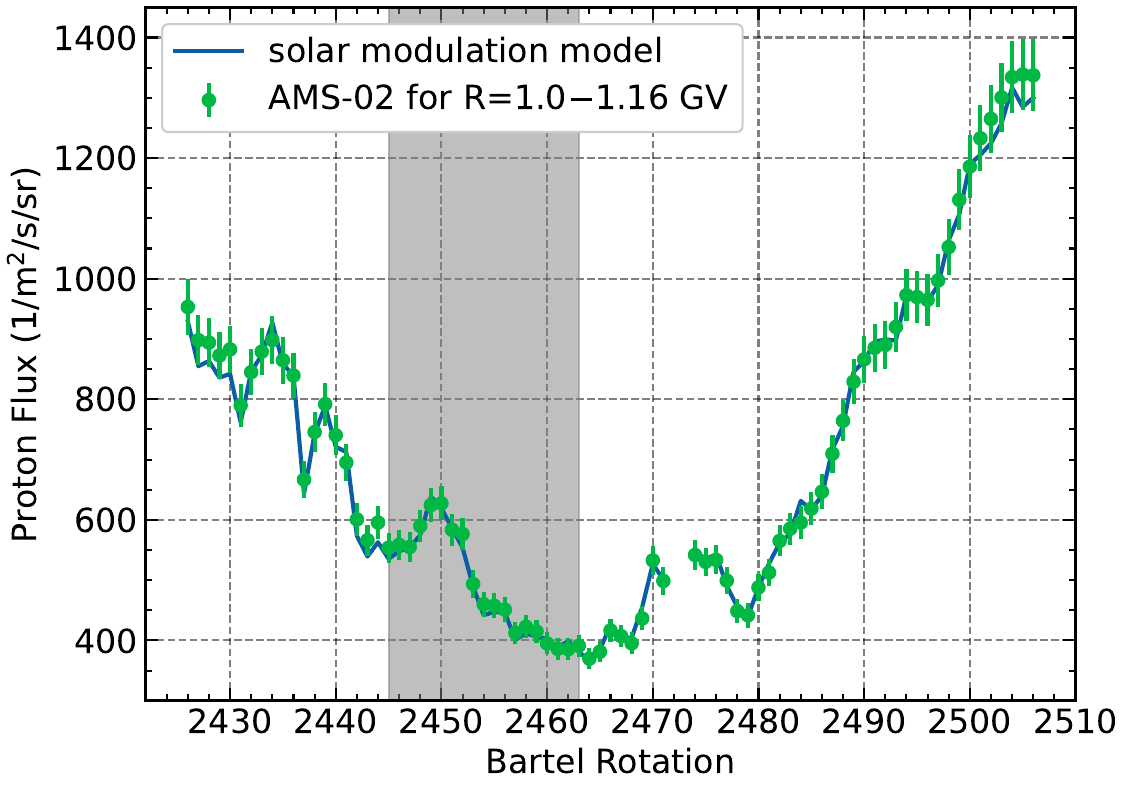}
    \includegraphics[width=0.3\textwidth]{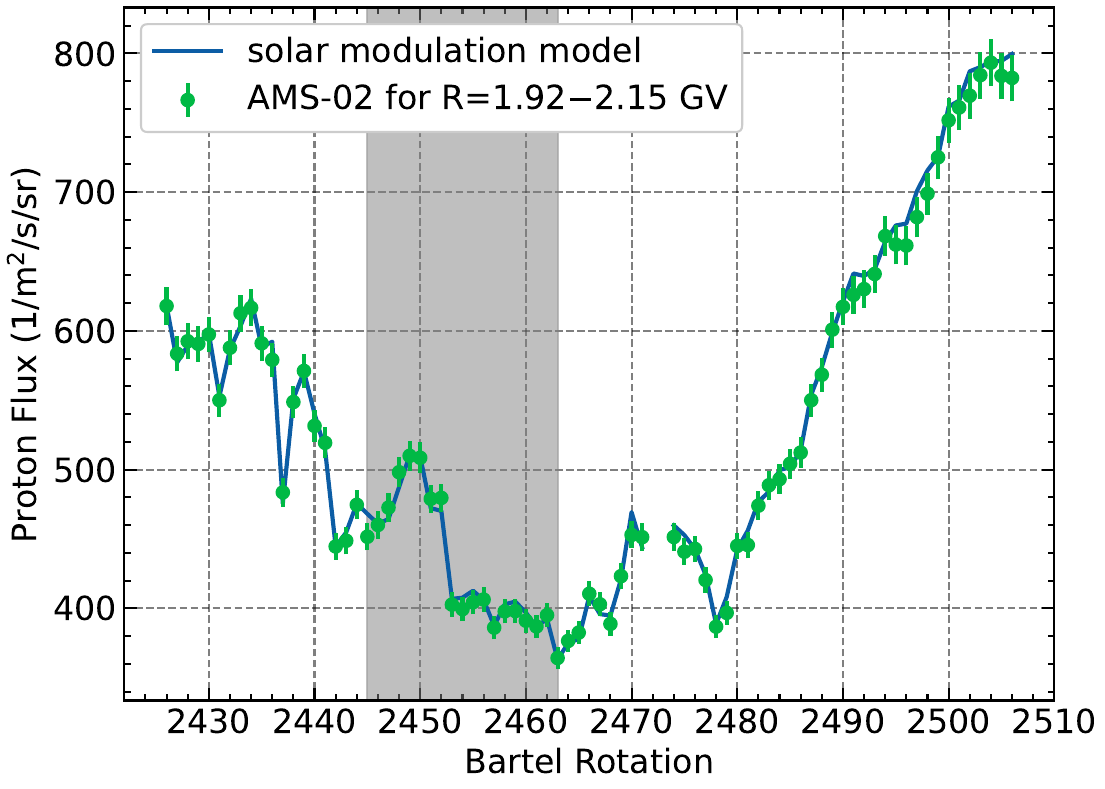}
    \includegraphics[width=0.3\textwidth]{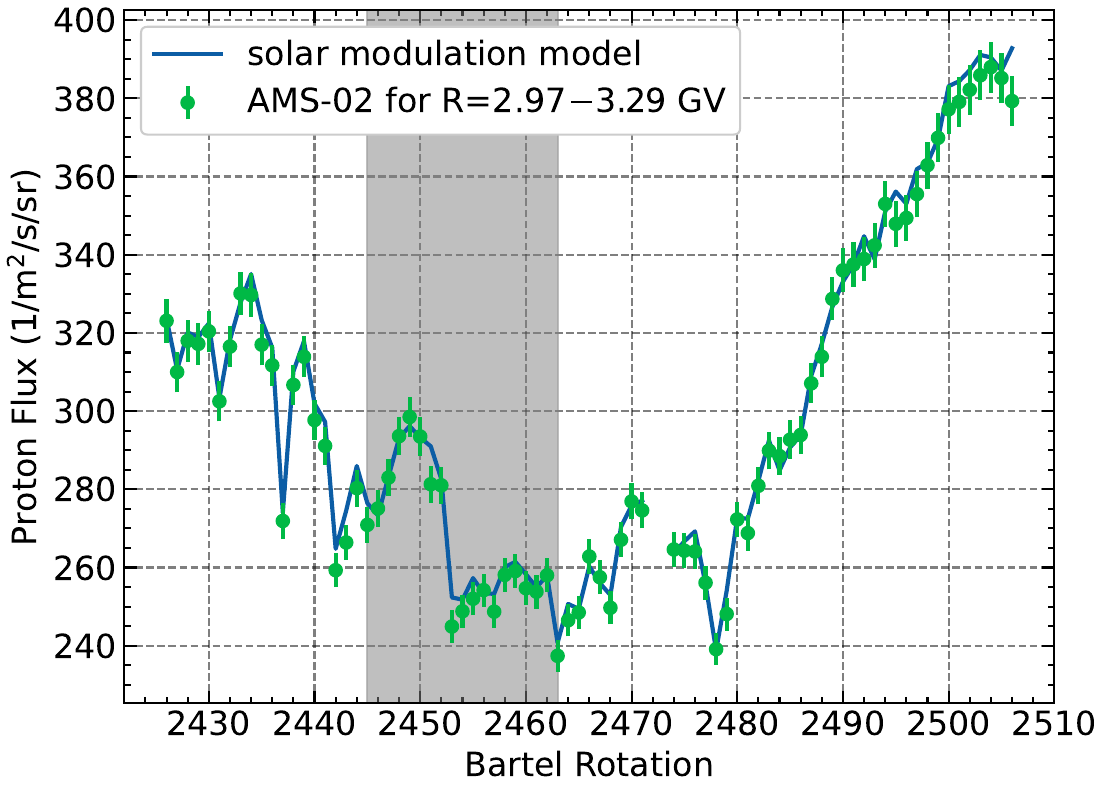}
    \includegraphics[width=0.3\textwidth]{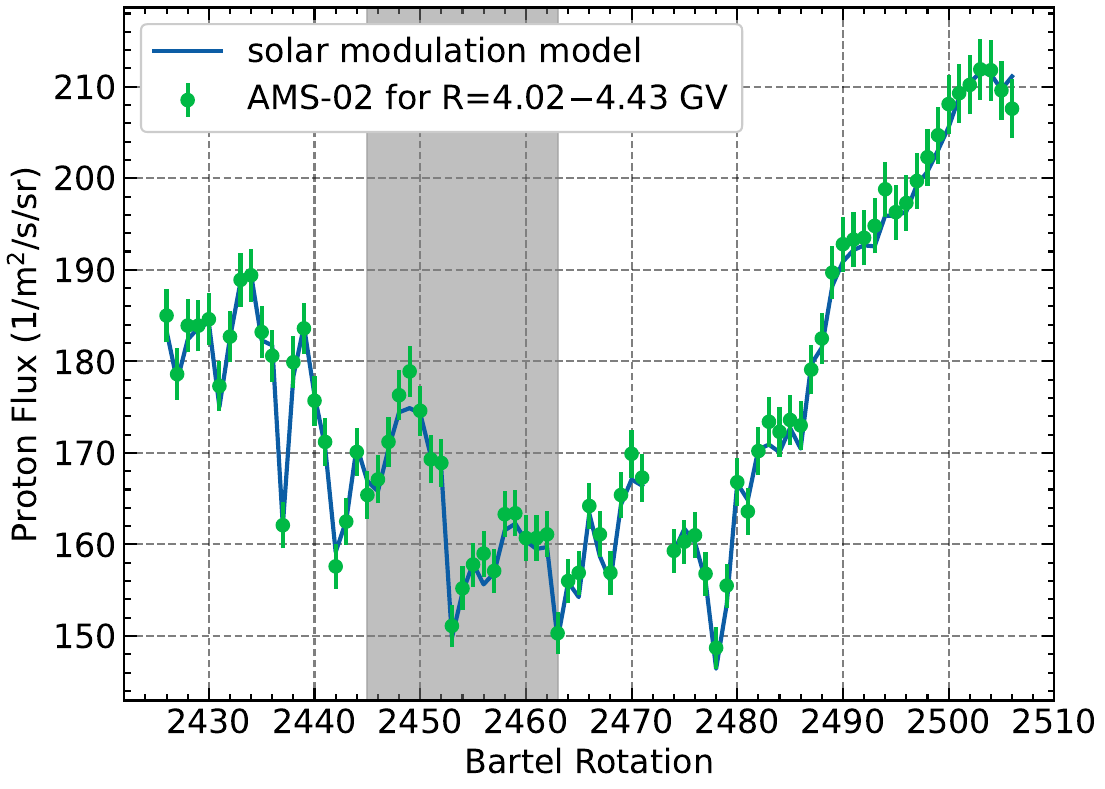}
    \includegraphics[width=0.3\textwidth]{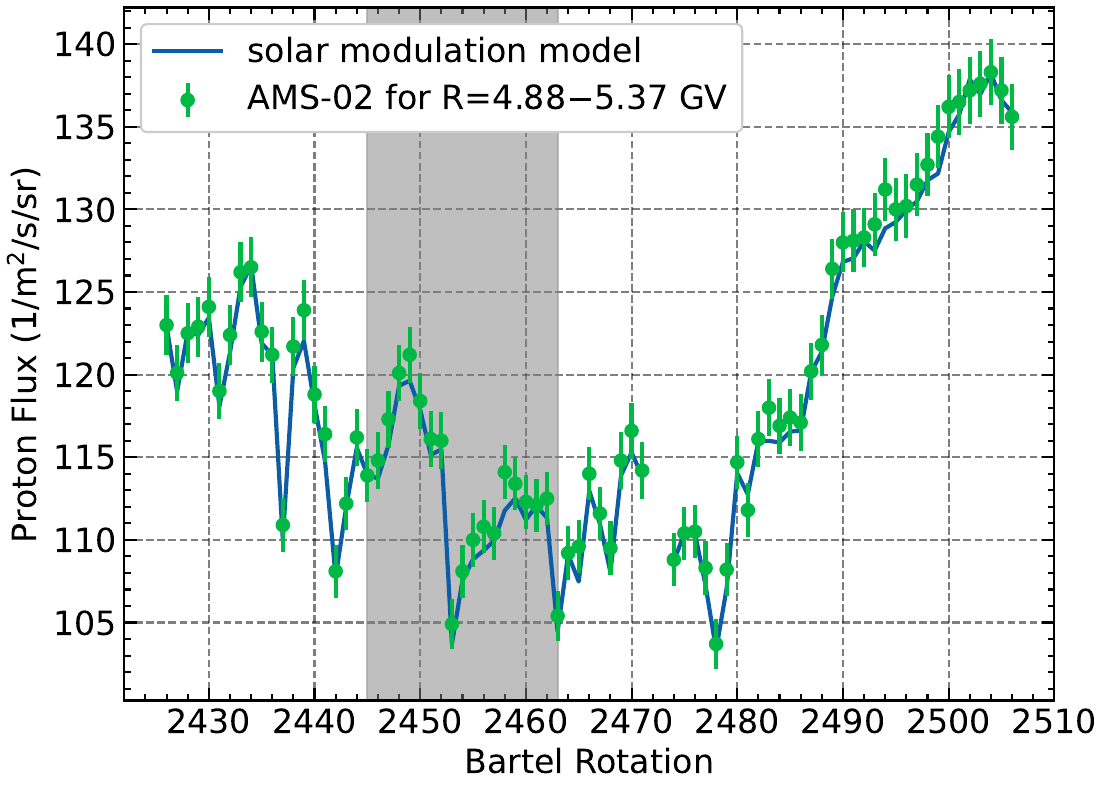}
    \includegraphics[width=0.3\textwidth]{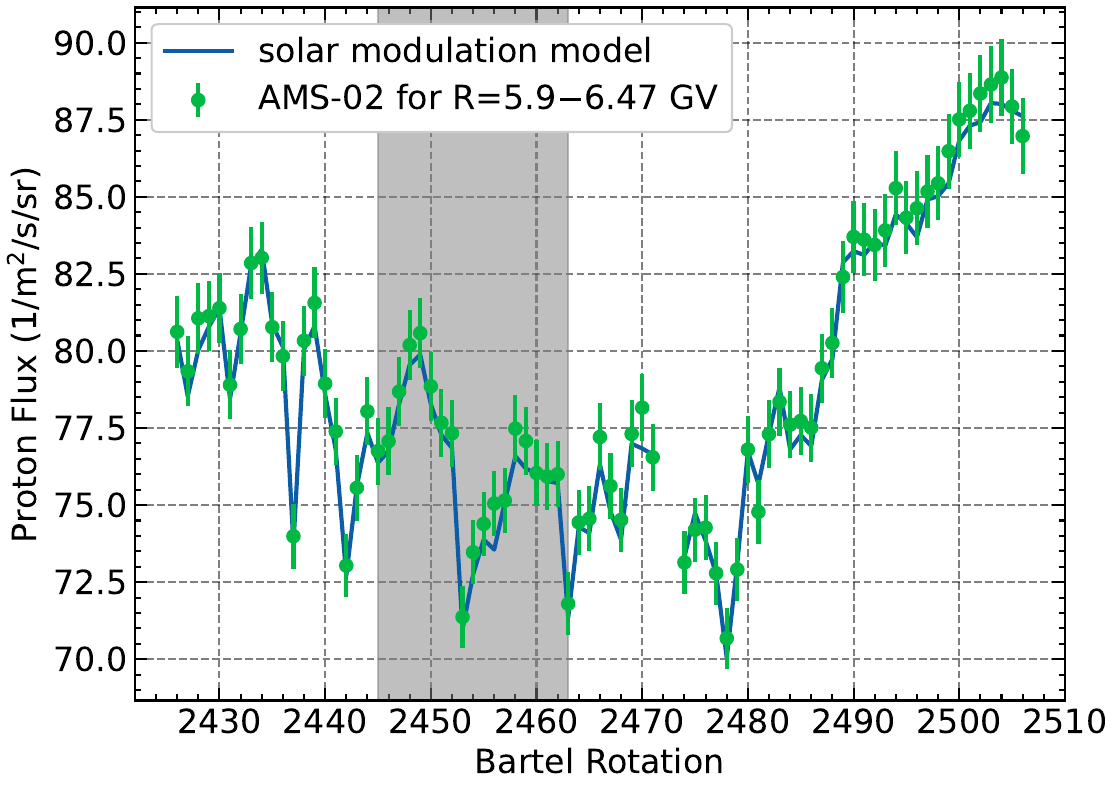}
    \includegraphics[width=0.3\textwidth]{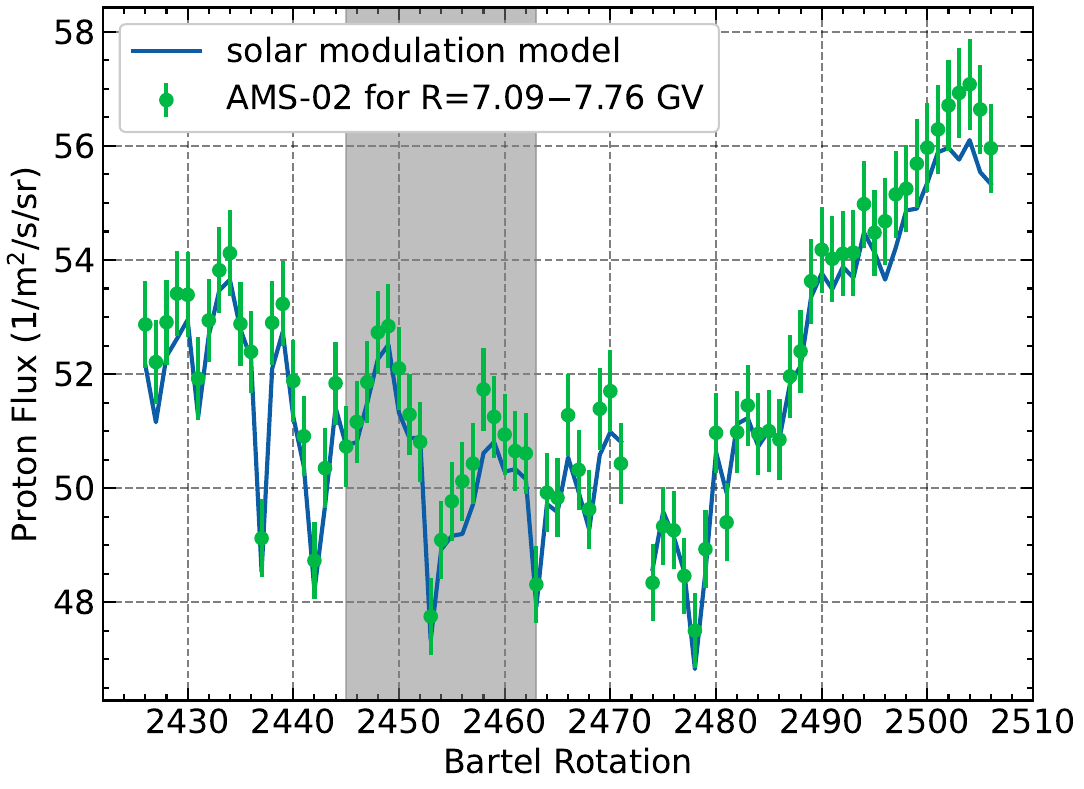}
    \includegraphics[width=0.3\textwidth]{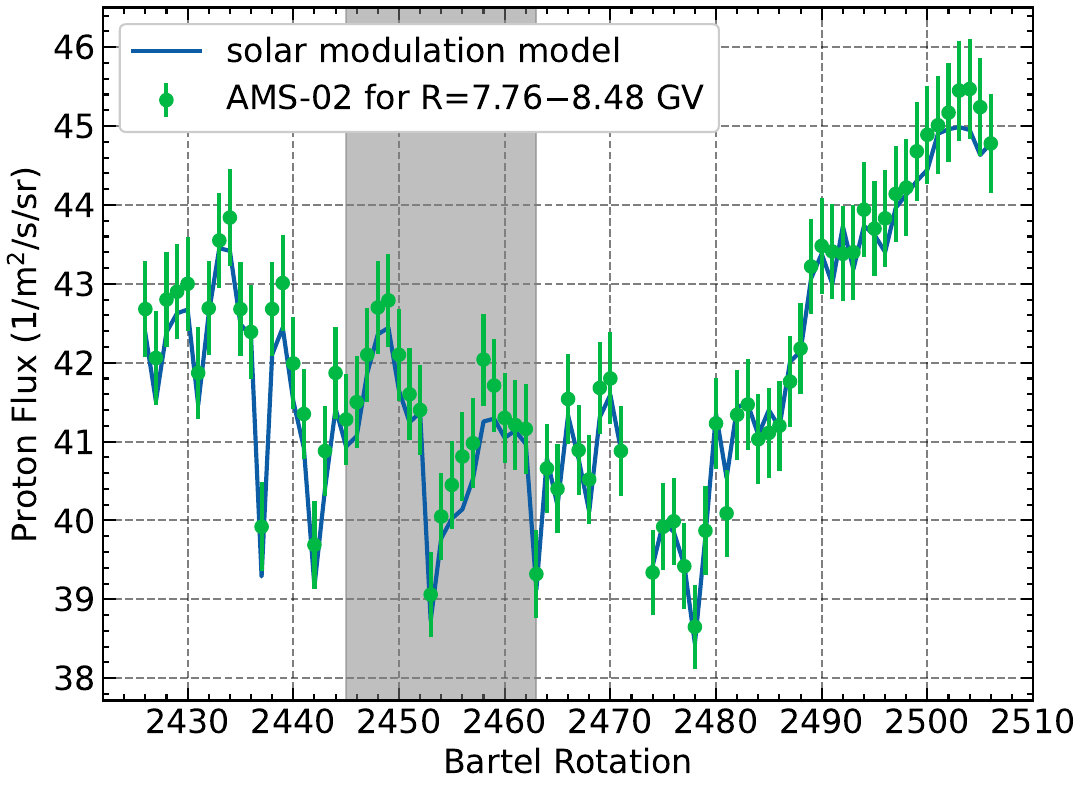}
    \includegraphics[width=0.3\textwidth]{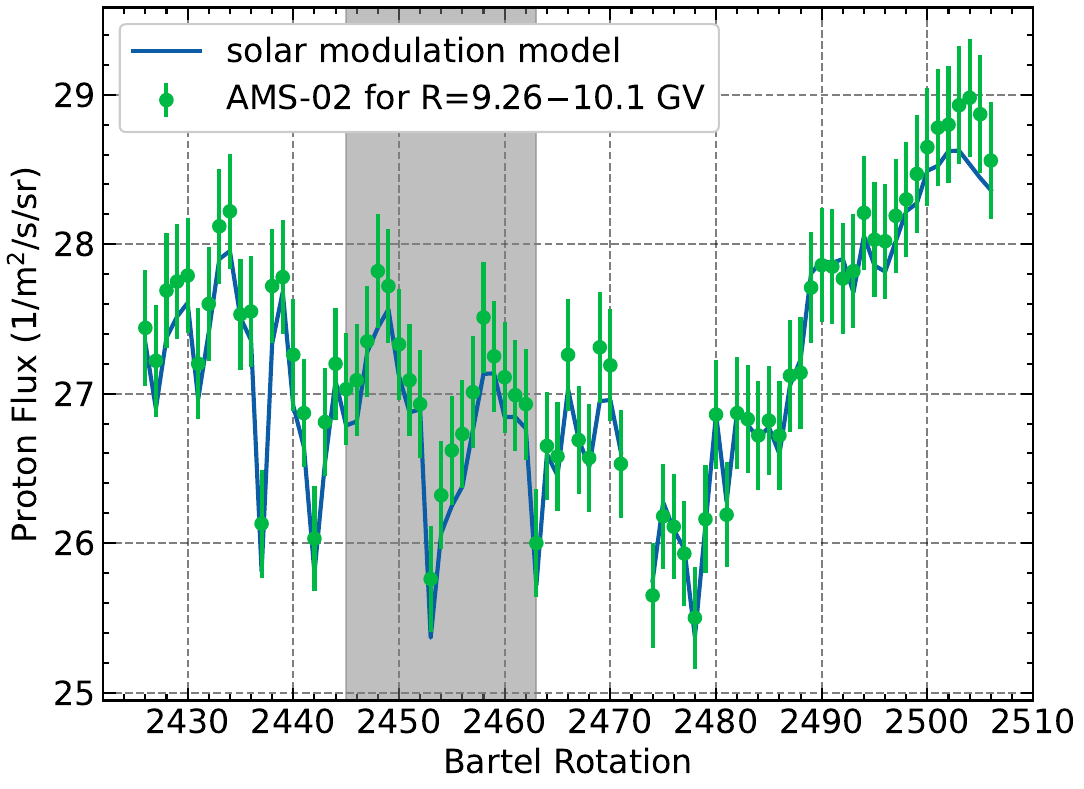}
    \caption{The time variation of the cosmic-ray proton flux across nine distinct ranges of rigidity, modulated by the BR cycles using the best-fit parameters, is compared with measurements from the AMS-02 experiment between May 2011 and May 2017 (BR cycles from 2426 to 2507).}
    \label{fig:Jtime_proton}
\end{figure}

We numerically solve the Parker transport equation using 3D simulation framework, optimizing the free parameters through comparison with observational data. Figure~\ref{fig:Jtime_proton} presents the temporal evolution of cosmic-ray proton fluxes across nine rigidity bins, derived using our best-fit modulation parameters. The modeled time variations, accounting for solar modulation through BR cycles 2426-2507 (from May 2011 to May 2017), show excellent agreement with AMS-02 measurements.

\bibliographystyle{JHEP}
\bibliography{biblio_dkk}
\end{document}